\documentclass[webpdf,modern,mediumone]{oup-authoring-template}

\graphicspath{{Fig/}}
\DeclareMathOperator*{\argmax}{arg\,max}  
\usepackage{natbib}
\usepackage{booktabs}
\usepackage{tabularx}
\usepackage{multirow}
\usepackage{makecell}
\usepackage{rotating}
\usepackage[table]{xcolor}
\definecolor{stripblue}{HTML}{e1eeff}
\graphicspath{ {./figures/} }

\theoremstyle{thmstyleone}%
\theoremstyle{thmstyletwo}%
\theoremstyle{thmstylethree}%

\begin{document}

\journaltitle{Journal Title Here}
\DOI{DOI added during production}
\copyrightyear{2026}
\pubyear{YEAR}
\vol{X}
\issue{x}
\access{Published: Date added during production}
\appnotes{Paper}

\firstpage{1}

\title[Short Article Title]{Data Shared Neighbourhood Selection for multi-condition network inference}

\author[1,2$\ast$]{Blanche Francheterre}
\author[1]{Ruben Colindres Zuehlke}
\author[3]{Vivian Viallon}
\author[1]{Marc Chadeau-Hyam}
\author[2]{Julien Chiquet}

\address[1]{\orgdiv{Department of Epidemiology and Biostatistics}, \orgname{School of Public Health, Imperial College London}, \orgaddress{\postcode{London}, \country{UK}}}
\address[2]{\orgdiv{Université Paris-Saclay}, \orgname{AgroParisTech INRAE, UMR MIA, SolsTIS team}, \orgaddress{\state{Paris}, \country{France}}}
\address[3]{\orgdiv{Nutrition and Metabolism Branch}, \orgname{International Agency for Research
on Cancer (IARC/WHO)}, \orgaddress{\state{Lyon}, \country{France}}}

\corresp[$\ast$]{Corresponding author. \href{email:email-id.com}{blanche.francheterre23@imperial.ac.uk}}

\received{Date}{0}{Year}
\revised{Date}{0}{Year}
\accepted{Date}{0}{Year}

\abstract{External stresses may affect both the circulating levels of specific biomarkers and disturb the correlation structures across molecular entities. The contribution of both types of dysregulations to the subsequent risk of disease are yet to be evaluated. We propose Data Shared Neighbourhood Selection (DSNS), a joint network inference method for estimating preserved and altered conditional association structures across related conditions. DSNS combines neighbourhood selection with the Data Shared Lasso decomposition, representing each nodewise regression coefficient as the sum of a shared component and a sparse condition-specific deviation. This provides an interpretable decomposition of molecular associations while retaining the computational advantages of neighbourhood selection. We also adapt the Stability Approach to Regularisation Selection (StARS) to this two-parameter joint estimation setting. In simulations involving sparse, hub-based and rewiring perturbation mechanisms, DSNS matched the best joint estimation methods for two conditions and outperformed them as the number of conditions increased, while remaining substantially faster than graphical lasso frameworks. Applied to prediagnostic inflammatory proteomic data from future lung cancer cases and matched controls in the EPIC-Italy and NOWAC cohorts, DSNS highlighted altered associations involving CDCP1 and IL10, two established lung cancer risk markers, as well as differential associations involving proteins not selected by risk models.} 


\maketitle


\section{Introduction}

Many biological processes arise from coordinated activity among molecular features rather than from isolated changes in individual markers. Proteins, genes and metabolites interact within complex systems, and comparisons across biological conditions may reveal differences not only in molecular levels, but also in the association structure linking them \citep{bassett2009human, chiquet2011inferring}. Such differences may reflect mechanistic contribution to the biological response to an external stress and/or to the development of a clinical outcome. Consequently, network-based molecular analyses can provide complementary insight to biomarker identification approaches by characterising how molecular features are connected. In disease studies involving related states, such as cases and controls, disease subtypes or treatment groups, a key objective is therefore not only to identify molecular markers associated with disease, but also to determine whether pairwise relationships across these markers are preserved or altered across biological states. Such analyses are particularly relevant in omics studies, where molecular relationships are expected to be shared across related conditions, with disease or treatment affecting only a subset of associations \citep{danaher2014joint}. Network analysis may help identifying pairwise correlations that may differ across conditions and contribute to the biological response of interest, and may therefore identify molecules that are involved in biolgoical response and pathogenesis that would not have been identified through their marginal effect.

A variety of computational methods have been proposed to infer such molecular association networks \citep{kuismin2017estimation}. Among them, Gaussian graphical models (GGMs) provide a natural framework for modelling interpretable biological networks and have become a widely used tool \citep{barabasi2004network}. The network structure is encoded by the precision matrix with zero off-diagonal entries correspond to conditionally independent pairs of variables and therefore to the absence of an edge in the graph. Compared with marginal correlation networks, GGMs distinguish direct conditional associations from associations induced indirectly through other variables. In omics studies, however, estimating GGMs is challenging because the number of molecular features is often comparable to, or larger than, the number of observations. To address this issue, a variety of regularised estimators have been proposed, including constrained $\ell_1$ minimisation approaches, Bayesian graphical models and regression-based methods. Among the most widely used are the graphical lasso, which estimates a sparse precision matrix directly through penalised likelihood \citep{friedman2008sparse}, and neighbourhood selection, which recovers graph structure through sparse nodewise regressions \citep{meinshausen2006high}. Unlike graphical lasso approaches, Neighbourhood Selection (NS) primarily estimates graph support rather than precision matrices, and is computationally efficient because the estimation problem decomposes into independent regressions.

When several biological states are observed, estimating each network independently ignores the substantial structure they are likely to share. Joint graphical model estimators leverage this shared structure by estimating several networks simultaneously while encouraging similarity across conditions through joint regularisation \citep{guo2011joint, chiquet2011inferring, danaher2014joint}. By borrowing information across related conditions, these methods improve graph recovery and differential edge detection relative to separate estimation \citep{danaher2014joint}. Existing approaches include the Joint Graphical Lasso \citep{danaher2014joint}, Joint Neighbourhood Selection \citep{chiquet2011inferring, zhang2012learning}, CLIME-based estimators \citep{lee2015joint} and Bayesian approaches \citep{li2019bayesian}; see \cite{tsai2022joint} for a review. 

Existing joint extimators typically estimate each condition's network jointly, with differential networks identified after estimation by comparing the resulting networks. Thus, shared and altered structures are encouraged through joint penalties but are not represented explicitly within the model. In many biological applications, however, it may be more natural to view their network structures as coming from a common underlying component with a limited number of alterations specific to each condition. Recent work on Ising models has shown the value of explicitly decomposing network parameters into shared and condition-specific components \citep{ballout2019structure}. Comparable formulations have received little attention for Gaussian graphical models.

We propose the Data Shared Neighbourhood Selection (DSNS), a joint network inference method which combines the Data Shared Lasso (DSL) coefficient decomposition with Neighbourhood Selection. Originally proposed for regression models, the Data Shared Lasso decomposes each condition's coefficients as the sum of a shared effect and a condition-specific deviation \citep{ollier2014joint, gross2016data}. We extend this principle to Neighbourhood Selection by decomposing each nodewise regression coefficient into a common component and deviations from that component for each condition. Consequently, DSNS simultaneously estimates the network structure shared across conditions and the sparse perturbations that distinguish them.  For each node, estimation reduces to a Data Shared Lasso problem that can be reformulated as a standard lasso estimation on an extended design matrix. Hence, this decomposition preserves the computational advantages of both Neighbourhood Selection and DSL.
A practical challenge is the calibration of the two regularisation parameters controlling the sparsity of the shared component and of the condition-specific deviations. We therefore adapt the Stability Approach to Regularisation Selection (StARS) \citep{liu2010stability} to the two-parameter joint estimation setting.

We evaluate DSNS against competing methods through an extensive simulation study reflecting key features of molecular networks, such as sparsity, degree heterogeneity and hub structure with various perturbation mechanisms. We apply DSNS to prediagnostic circulating inflammatory protein measurements from the study of \citet{dagnino2021prospective}. By jointly estimating the protein association networks of future lung cancer cases and matched controls, we aim to identify associations that are preserved across groups and those that differ before diagnosis, thereby providing complementary insight into the inflammatory processes associated with future lung cancer risk.

\section{Methods}
\subsection{Background}
Consider $K$ datasets coming from different biological conditions, for example controls and cases or different disease subtypes. For each condition $k \in \{1, \ldots, K\}$, we measure $p$ molecular markers (e.g. proteomics, gene expression, metabolites) for $n_k$ independent individuals. We assume that the observations, $X_{\ell}^{(k)}$, follow a multivariate Normal distribution such that: 
$$X_{\ell}^{(k)} \sim \mathcal{N}(\mu^{(k)}, \Sigma^{(k)}),\quad \ell \in \{1,\ldots, n_k\}, \  k \in \{1,\ldots, K\}$$ where $\mu^{(k)} \in \mathbb{R}^p$ is the mean vector and $\Sigma^{(k)} \in \mathbb{R}^{p \times p}$ the covariance matrix for condition $k$. Throughout the paper, we assume that variables are centered within each condition. For each condition, the precision matrix is defined as the inverse of the covariance matrix, $\Omega^{(k)} = \Sigma^{(k)^{-1}}$, and is assumed to be sparse. Specifically, under the Gaussian graphical model assumption, $\Omega_{ij}^{(k)} = 0, i\neq j$ if and only if $X_i^{(k)}$ and $X_j^{(k)}$ are independent conditionally on all other variables. Each condition's precision matrix encodes an undirected graph $\mathcal{G}^{(k)} = (V, \mathcal{E}^{(k)})$, where 
$V = \{1, \ldots, p\}$ is the set of  shared molecular features and the edge set is:
$$\mathcal{E}^{(k)} = \{(i,j) : \Omega_{ij}^{(k)} \neq 0, \ i \neq j \} $$
We consider two objectives: first, recovery of the graph supports for each biological state, by estimating $\{\mathcal{E}^{(k)}\}_{k=1}^K$ with the estimation of the precision matrices $\{\Omega^{(k)}\}_{k=1}^K$ and second, identification of differential biological networks across conditions. 

Two complementary definitions of differential network are considered \citep{shojaie2021differential}. The first is differential support, also referred to as differential structure, which captures differences in edge across conditions. 
For $K$ conditions, we define the differential support as the set of edges that are present in at least one condition and absent in at least one other condition:
$$
\mathcal{E}_{\mathrm{suppdiff}}
=
\left(\bigcup_{k=1}^{K} \mathcal{E}^{(k)}\right)
\setminus
\left(\bigcap_{k=1}^{K} \mathcal{E}^{(k)}\right).
$$

The second is differential edge weights which capture changes in the strength of conditional dependencies, including changes that occur without support differences. It is defined pairwise between conditions as: $$\mathcal{E}_{\text{precdiff}}^{(kk')} = \{(i,j) : \Omega_{ij}^{(k)} \neq \Omega_{ij}^{(k')} , \ i \neq j \}, k \neq k'.$$

We introduce the two main families of joint inference methods: Joint Graphical Lasso approaches, which directly estimate the precision matrix via penalised likelihood, and Joint Neighbourhood Selection, which estimate graph structure via nodewise regressions. Both families introduce a joint penalty that encourages similarities across conditions and involve two non-negative regularisation parameters: 
$\lambda_1$, which controls within-condition sparsity, and $\lambda_2$, controlling the degree of similarity enforced across conditions. 

\paragraph*{Joint Graphical Lasso} \label{sec_jgl}
The Joint Graphical Lasso (JGL), introduced by \cite{danaher2014joint} , extends the standard graphical Lasso to multiple conditions and estimates $\{\Omega^{(k)}\}_{k=1}^K$ by minimising the penalised log-likelihood:
$$
\min_{\{\Omega^{(k)}\}_{k=1}^K}
\sum_{k=1}^{K} n_k \left[ \operatorname{tr}\!\left( \hat\Sigma^{(k)} \Omega^{(k)} \right)- \log{\det(\Omega^{(k)})} \right] + 
\lambda_1 \sum_{k=1}^K \sum_{i\neq j} \left| \Omega^{(k)}_{ij} \right| + 
\lambda_2 \mathcal{P}(\{\Omega^{(k)}\})
$$
where $\hat\Sigma^{(k)}$ is the empirical covariance matrix for condition $k$ and  $\mathcal{P}$ a joint penalty across conditions. The estimated precision matrices returned by JGL are sparse, symmetric and positive-definite.

Two commonly used penalties are the Fused Graphical Lasso (FGL), which encourages similar edge weights across conditions, and the Group Graphical Lasso (GGL), which encourages common sparsity patterns. Additional variants such as Perturbed-node Joint Graphical Lasso (PNJGL) \citep{mohan2014node} assume that differences are concentrated around a subset of perturbed nodes. Details are provided in Supplementary Material \ref{app_jgl}.

\paragraph*{Joint Neighbourhood Selection}\label{sec_ns}

Neighbourhood Selection (NS), introduced by \cite{meinshausen2006high}, estimates the conditional dependency structure by estimating $p$ penalised regressions where each node $j$ is regressed against all other variables. Under the Gaussian graphical model assumption, the regression coefficients satisfy: $$\beta_{ij} = -\frac{\Omega_{ij}}{\Omega_{jj}}$$ where $\beta_{ij}$ denotes the coefficient associated with predictor $i\neq j$ in the regression of node $j$. Thus, nonzero coefficients correspond to nonzero off-diagonal entries in the precision matrix and therefore to edges in the graph.

Similarly to the graphical lasso, Neighbourhood Selection can be extended to multiple conditions by introducing a joint penalty $\mathcal{P}$ between regression coefficients across conditions \citep{chiquet2011inferring, zhang2012learning}. For each node $j$, we solve:
$$
\min_{\{\beta_j^{(k)}\}_{k=1}^K}\sum_{k=1}^K\frac{1}{2n_k}
\left\|X^{(k)}_j - X_{-j}^{(k)} \beta^{(k)}_{j}\right\|_2^2+
\lambda_1\sum_{k=1}^K\left\| \beta^{(k)}_{j} \right\|_1 + \lambda_2\mathcal{P}(\{\beta_j^{(k)}\})
$$
where $X_{j}^{(k)} \in \mathbb{R}^{n_k}$ is the $j$th column of $X^{(k)}$, $X_{-j}^{(k)} \in \mathbb{R}^{n_k \times (p-1)}$ consists of all variables of $X^{(k)}$ except $j$ and $\beta_j^{(k)} \in \mathbb{R}^{p-1}$, the coefficient vector for node $j$.

Analogously to JGL, two common penalties are Fused Neighbourhood Selection (FNS) \citep{zhang2012learning}, which encourages similar coefficient values across conditions, and Group Neighbourhood Selection (GNS) \citep{chiquet2011inferring, liu2025learning}, which encourages common sparsity patterns. Additional details are provided in Supplementary Material \ref{app_jns}.

As the $p$ regressions are solved independently for each node, the resulting edge set is not necessarily symmetric. Symmetry is enforced by post-processing using either the AND rule, an edge $(i,j)$ is included in the graph if $\hat\beta_{ij} \neq 0$ and $\hat\beta_{ji} \neq 0$ or the OR rule, an edge $(i,j)$ is included in the graph if $\hat\beta_{ij} \neq 0$ or $\hat\beta_{ji} \neq 0$. We write $\hat{S}^{\text{AND}}(\hat{\beta})$
and $\hat{S}^{\text{OR}}(\hat{\beta})$ for the supports obtained
by applying these rules to the estimated $\hat{\beta}$.

To ensure comparable implementation across methods, all neighbourhood selection approaches considered in this work were implemented within our R package \href{https://forge.inrae.fr/blanche.francheterre/monique}{\textit{monique}}.

\subsection{Data Shared Neighbourhood Selection model}
We consider a decomposition in which each condition's network is represented as the sum of a shared component and a sparse deviation, allowing common and differential network structures to be estimated simultaneously. The \textit{Data Shared Neighbourhood Selection} (DSNS) is a joint network inference method that combines the coefficient decomposition strategy of Data Shared Lasso with Neighbourhood Selection.

\paragraph*{Data Shared Lasso} \label{sec_dsl}

The Data Shared Lasso (DSL), introduced by \cite{ollier2014joint, gross2016data}, was developed for regression settings in which observations are partitioned into $K$ non-overlapping conditions defined by a categorical covariate. 
The DSL assumes that the response is generated as:
$$y_i = x_i^T(\boldsymbol{\theta} + \boldsymbol{\Delta^{(k_i)}})+ \epsilon_i$$

where $k_i \in \{1, \ldots, K\}$ denotes the condition of observation $i$, $\boldsymbol{\theta} \in \mathbb{R}^p$ is the coefficient vector shared across all conditions, $\boldsymbol{\Delta^{(k)}} \in \mathbb{R}^p$ is the deviation for each condition $k$, and $\epsilon_i$ are independent errors.

Both components are estimated jointly by minimising:  
$$
\min_{\theta, \{\Delta^{(k)}\}_{k=1}^K}\sum_{i=1}^n\frac{1}{2n}
\left(y_i - x_i^T(\boldsymbol{\theta} + \boldsymbol{\Delta^{(k_i)}})\right)^2+
\lambda \left( \left\| \boldsymbol{\theta} \right\|_1+ \sum_{k=1}^K\left\| \boldsymbol{\Delta^{(k)}} \right\|_1
\right)
$$
where $\lambda$ is the regularisation parameter. 
The authors showed that estimation of these components can be performed as a standard lasso problem through a simple redesign of the predictor matrix, making the approach computationally efficient.

In the following section, we integrate this decomposition into the NS framework to jointly estimate multiple related graphs.

\paragraph*{Model} \label{sec_dsns}

\begin{figure}[h]%
\centering
\includegraphics[width=0.99\textwidth]{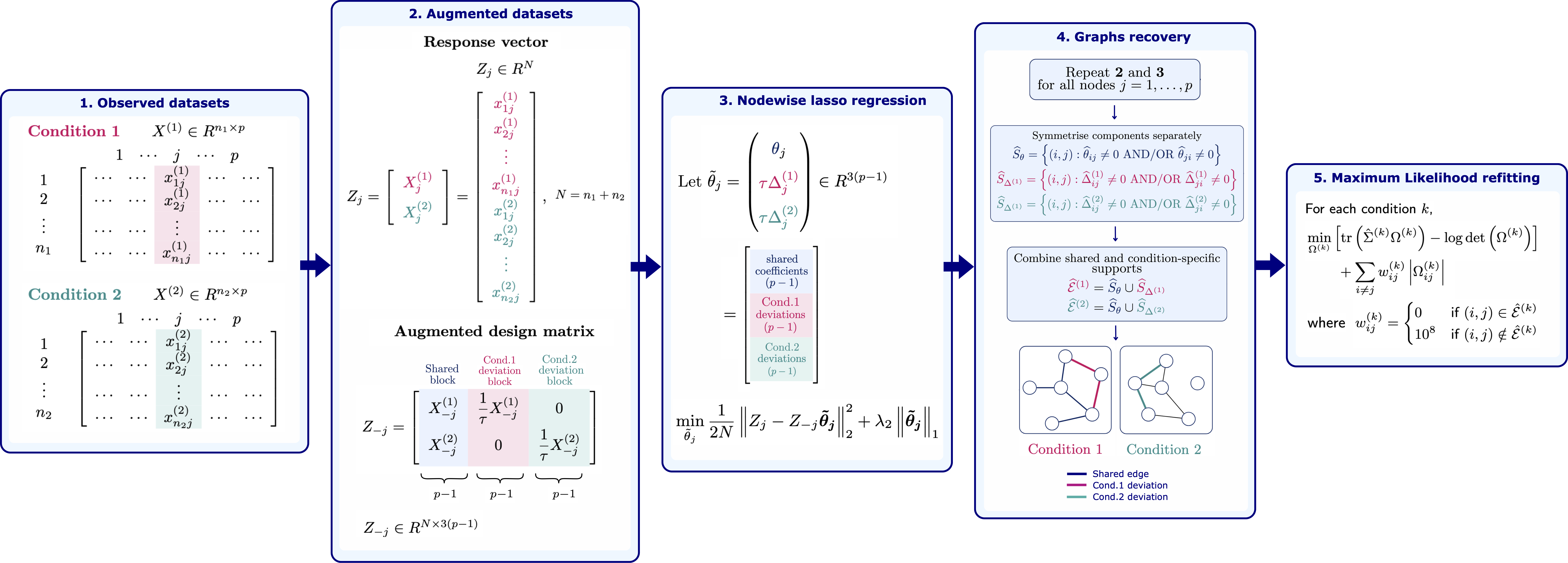}
\caption{DSNS framework under $K=2$.}\label{fig_dsns}
\figalttext[Schematic flow diagram of DSNS framework in four steps, left to right.]{Schematic flow diagram in four stages, left to right. Step 1 shows two observed data matrices, one per condition, of dimension $n_1 \times p$ and $n_2 \times p$, with column $j$ highlighted. Step 2 shows these stacked into a single response vector $Z_j$ of length N, and an extended design matrix $Z_{-j}$ with three column blocks of width $p−1$: a shared block containing both conditions' predictors, and two condition-specific blocks scaled by $1/\tau$, each zero outside its own condition's rows. Step 3 shows the resulting coefficient vector partitioned into shared coefficients and two deviation blocks, estimated by a single lasso problem. Step 4 shows the shared and deviation supports symmetrised separately, then combined by union into two condition-specific graphs, drawn with shared edges in one colour and condition-specific deviations in two others.}%
\end{figure}

Motivated by the assumption that related conditions share a common network structure, DSNS assumes  that each condition node's neighbourhood is the sum of a sparse shared component and a sparse deviation specific to each condition. For each node $j$ and condition $k$, we have:
$$\boldsymbol{\beta_j^{(k)}} = \boldsymbol{\theta_j} + \boldsymbol{\Delta_j^{(k)}}$$
where $\boldsymbol{\beta_{j}^{(k)}} \in \mathbb{R}^{p-1}$ denotes the vector of regression coefficients, $\boldsymbol{\theta_{j}} \in \mathbb{R}^{p-1}$ represents the coefficients shared across conditions and $\boldsymbol{\Delta_{j}^{(k)}} \in \mathbb{R}^{p-1}$ represents the condition's deviation.

We assume that each node $j$ is generated as:
$$X_j^{(k)} = X_{-j}^{(k)} \boldsymbol{\beta_j^{(k)}} + \boldsymbol{\epsilon} = X_{-j}^{(k)} (\boldsymbol{\theta_j} + \boldsymbol{\Delta_j^{(k)}})+ \boldsymbol{\epsilon}$$
Thus, for each node $j$, DSNS estimates the coefficient components by solving the following minimisation:
\begin{equation}
\min_{\boldsymbol{\theta_j}, \{\boldsymbol{\Delta_j^{(k)}}\}_{k=1}^K}\sum_{k=1}^K\frac{1}{2N}
\left\|X_j^{(k)} - X_{-j}^{(k)}(\boldsymbol{\theta_{j}} + \boldsymbol{\Delta_{j}^{(k)}})\right\|_2^2+
\lambda_1\sum_{k=1}^K\left\| \boldsymbol{\Delta_{j}^{(k)}} \right\|_1+
\lambda_2\left\| \boldsymbol{\theta_{j}} \right\|_1
\label{eqdsns}
\end{equation}
where the penalty parameter $\lambda_1$ controls the sparsity of each condition's deviations while $\lambda_2$ controls the sparsity of the shared component and $ N = \sum_{k=1}^K n_k$. Although the decomposition is not identifiable in the absence of regularisation, the $\ell_1$ penalty provides an optimal sparse decomposition under the conditions described in Supplementary Material \ref{app_dsns_iden} \citep{ballout2019structure}.

For each node $j$, estimating the shared and deviation components corresponds to solving a DSL problem. Consequently, the optimisation can be reformulated as a standard lasso problem through an extended design matrix. For each node $j$, we define the response vector: \begin{equation*} Z_{j}= \begin{pmatrix}
X_{j}^{(1)} \\
X_{j}^{(2)} \\
\vdots  \\
X_{j}^{(K)} \\ 
\end{pmatrix} \in \mathbb{R}^{N} \end{equation*}
and construct the extended design matrix:
\begin{equation*}
Z_{-j}= \begin{pmatrix}
X_{-j}^{(1)} & \frac{1}{\tau} X_{-j}^{(1)} & 0 & \cdots & 0  \\
X_{-j}^{(2)} & 0 & \frac{1}{\tau} X_{-j}^{(2)} & \cdots & 0  \\
\vdots & \vdots & \vdots & \ddots & \vdots \\
X_{-j}^{(K)} & 0 & 0 & \cdots & \frac{1}{\tau} X_{-j}^{(K)} \\
\end{pmatrix} 
\in \mathbb{R}^{N \times (p-1)(K+1)}
\end{equation*}
where the first block of $p-1$ columns corresponds to the shared predictors while the remaining $K$ blocks encode the predictors specific to each condition and $\tau = \lambda_1 / \lambda_2$ is the ratio of penalty parameters.

Let $\boldsymbol{\tilde\theta_j} = (\boldsymbol{\theta_j},\tau \boldsymbol{\Delta_j^{(1)}},\dots, \tau \boldsymbol{\Delta_j^{(K)}})^\top,$ then
\autoref{eqdsns} can be rewritten as:
$$
\min_{\tilde\theta_j}\frac{1}{2N}
\left\|Z_j - Z_{-j}\boldsymbol{\tilde\theta_{j}}\right\|_2^2+
\lambda_2 \left\| \boldsymbol{\tilde\theta_{j}} \right\|_1
$$
Similarly to the DSL, this reformulation shows that DSNS can be estimated using existing lasso solvers without requiring specialised optimisation algorithms \citep{tibshirani1996regression}.

\paragraph*{Symmetrisation}
 As in standard neighbourhood selection, edge symmetry needs to be enforced as a post-processing step but for DSNS, we symmetrise the shared and condition-specific components separately to preserve their individual interpretability. Applying the AND and OR rules to the shared component
$\hat{\boldsymbol{\theta}}$ and to each condition-specific component
$\hat{\boldsymbol{\Delta}}^{(k)}$ gives the supports
$\hat{S}^{\text{AND}}_{\theta}, \hat{S}^{\text{OR}}_{\theta}$ and
$\hat{S}^{\text{AND}}_{\Delta^{(k)}}, \hat{S}^{\text{OR}}_{\Delta^{(k)}}$. As the two components are symmetrised independently, any pairing of their
rules can be considered, giving four combinations:
$(\hat{S}^{\text{AND}}_{\theta}, \hat{S}^{\text{AND}}_{\Delta^{(k)}})$,
$(\hat{S}^{\text{OR}}_{\theta}, \hat{S}^{\text{OR}}_{\Delta^{(k)}})$,
$(\hat{S}^{\text{AND}}_{\theta}, \hat{S}^{\text{OR}}_{\Delta^{(k)}})$ and
$(\hat{S}^{\text{OR}}_{\theta}, \hat{S}^{\text{AND}}_{\Delta^{(k)}})$.

For each condition $k$, the estimated edge set is then obtained by combining the
shared and condition-specific supports,
\[
\hat{\mathcal{E}}^{(k)}
  = \hat{S}_{\theta} \cup \hat{S}_{\Delta^{(k)}},
\]
where each support is symmetrised according to the chosen combination. While joint estimators presented above recover differential networks by comparing estimated supports after fitting, the DSNS parameterisation gives the differential structure directly: $\hat{S}_{\Delta^{(k)}}$, symmetrised by the chosen rule.

The framework is shown for $K=2$ in \autoref{fig_dsns}.

\paragraph*{Maximum Likelihood refitting}
Neighbourhood Selection methods primarily aim at recovering the graph support rather than accurate estimation of precision matrices. Thus, to obtain these estimates, a refitting step is performed following support identification \citep{michailidis2016}. A graphical lasso model constrained to the inferred support is fitted in which entries corresponding to selected edges are left unpenalised, whereas excluded edges are penalised with a large constant penalty, $10^8$.
For each condition $k$, the precision matrix is estimated by solving:
\begin{equation}
\min_{\Omega^{(k)}}
\left[
\operatorname{tr}\left(\hat{\Sigma}^{(k)}\Omega^{(k)}\right) - \log\det\left(\Omega^{(k)}\right)
\right] + \sum_{i \neq j} w^{(k)}_{ij} \left|\Omega^{(k)}_{ij}\right|
\label{eqrefit}
\end{equation}
where:
\begin{equation*}
w^{(k)}_{ij} =
\begin{cases}
0 & \text{if } (i,j) \in \hat{\mathcal{E}}^{(k)} \\
10^8 & \text{if } (i,j) \notin \hat{\mathcal{E}}^{(k)}
\end{cases}
\end{equation*}
and $\hat{\mathcal{E}}^{(k)}$ is the support estimated by DSNS. 

This refitting procedure results in a symmetric positive-definite precision matrix estimate while preserving the support estimated by DSNS.

\subsection{Calibration}

Calibrating the two hyperparameters $(\lambda_1, \lambda_2)$ involves balancing sparsity, goodness-of-fit and stability of the inferred graphs. We consider two calibration strategies: information criteria and subsampling-based stability.

\paragraph*{Information criteria}

Well established information criteria such as Bayesian (BIC) or Akaike (AIC) Information Criterion \citep{foygel2010extended} evaluate the likelihood of the observed data under the estimated model parameters while penalising model complexity according to the number of estimated parameters. In high dimensional settings AIC and BIC tend to select overly dense graphs, leading to a high number of false positive edges. To address this issue, we consider the extended Bayesian Information Criterion (eBIC) which introduces an additional parameter, $\gamma$, that penalizes graph density and encourages sparser networks \citep{foygel2010extended}. For multiple conditions, the eBIC criterion at a given pair of sparsity parameters $(\lambda_1, \lambda_2)$ is defined as:
$$
  \mathrm{eBIC}(\lambda_1,\lambda_2) = \sum_{k=1}^K \left[-2\ell(\hat{\Omega}^{(k)}(\lambda_1,\lambda_2)) + |\mathcal{\hat{E}}^{(k)}(\lambda_1,\lambda_2)|\log(n_k) + 4\gamma |\mathcal{\hat{E}}^{(k)}(\lambda_1,\lambda_2)|\log(p) \right],
$$
where, for condition $k$, $\hat{\Omega}^{(k)}(\lambda_1,\lambda_2)$ denotes the refitted precision matrix, $|\mathcal{\hat{E}}^{(k)}(\lambda_1,\lambda_2)|$ the number of edges, $n_k$ the sample size and $p$ the number of variables.

\paragraph*{Subsampling-based calibration}

Information  criteria can give unstable results, especially in high-dimensional settings, where the same sparsity parameters or small changes in the data can give rise to different sets of selected edges \citep{liu2010stability}. To address this, we adapted StARS, which has been shown to be robust for single Gaussian graphical network inference calibration \citep{liu2010stability}, to the joint inference setting which involves two regularisation parameters.

For each condition, $B$ subsamples of size $n_k/2$ are drawn without replacement from the initial dataset. For each hyperparameter pair $(\lambda_{1}, \lambda_{2})$ on a two-dimensional grid, where larger $\lambda_{1}$ or $\lambda_{2}$ implies more regularisation, DSNS is fitted on each subsample and the selected edges are recorded for each condition.

Let $\Pi^{(k)}_{ij}(\lambda_1, \lambda_2)$ denote the proportion of subsamples in which edge $(i,j)$ is selected in condition $k$ at parameter $(\lambda_1, \lambda_2)$. The total instability at $(\lambda_1, \lambda_2)$ is defined as:
$$
  \hat D (\lambda_1, \lambda_2) = \frac{1}{K\cdot \frac{p (p-1)}{2}} \sum_{k=1}^K\sum_{i<j} 2\cdot \Pi_{ij}^{(k)}(\lambda_1, \lambda_2)(1-\Pi^{(k)}_{ij}(\lambda_1, \lambda_2))
$$
To ensure monotonicity of the instability measure with increasing regularisation, let:
$$
  \bar D (\lambda_1, \lambda_2) = \text{sup}_{\substack{t1 \geq \lambda_1 \\ t2 \geq \lambda_2}}\hat D(t_1, t_2)
$$
The regularisation parameters are chosen such that: $$(\hat{\lambda}_1, \hat{\lambda}_2)_{\text{StARS}} = \argmax_{(\lambda_1, \lambda_2)}| \hat{\mathcal{E}}(\lambda_1, \lambda_2)|, \text{ subject to } \bar D(\lambda_1, \lambda_2)\leq \beta \text{ and } | \hat{\mathcal{E}}(\lambda_1, \lambda_2)| > 0$$ where $| \hat{\mathcal{E}}(\lambda_1, \lambda_2)|$ denotes the total number of edges summed across conditions and $\beta$ is the instability threshold. Following the original article recommendation, we considered $\beta=0.05$ as the default value. In the simulation study, we also evaluated $\beta=0.10$ as a less conservative alternative. Finally, the retained graph is obtained by fitting the model on the full dataset at calibrated sparsity parameters $(\hat{\lambda}_1, \hat{\lambda}_2)_{\text{StARS}}$.

\subsection{Simulation study}

We evaluated the performance of DSNS against 9 competing methods with respect to overall graph recovery, differential support recovery, and differential precision recovery across simulation scenarios designed to reflect biologically relevant network configurations. The methods consisted of Joint Graphical Lasso methods: Fused Graphical Lasso (FGL), Group Graphical Lasso (GGL) and Perturbed-node Joint Graphical Lasso (PNJGL), and Joint Neighbourhood Selection methods: Fused Neighbourhood Selection (FNS) and Group Neighbourhood Selection (GNS). We considered their pooled and independent variants as baselines. Pooled network inference consists in inferring one graph on all samples regardless of condition while in the independent variant, each condition is estimated independently. For clarity, we reported the neighbourhood
selection methods under a single symmetrisation rule throughout: the (OR, OR) rule for DSNS and the OR rule for the neighbourhood selection methods. A comparison across the alternative rules is provided in the Supplementary \autoref{rules_dsns}, \ref{rules_ns}. Methods and joint penalties are summarised in \autoref{table1}. As available software implementations differ substantially across joint graphical model estimators, particular care was taken to harmonise the implementation of all nodewise methods in our R package \href{https://forge.inrae.fr/blanche.francheterre/monique}{\textit{monique}}. Additionally, to ensure fair comparison, the refitting step (\ref{eqrefit}) was performed for all joint estimation methods.

\begin{sidewaystable}
\centering
\renewcommand{\arraystretch}{2}
\setlength{\aboverulesep}{0pt}
\setlength{\belowrulesep}{0pt} 
\setlength{\extrarowheight}{0.4ex} 
\setlength{\tabcolsep}{8pt}
\begin{tabularx}{\textwidth}{p{2.2cm} p{6.1cm} >{\raggedright\arraybackslash}p{3cm} X}
\toprule
\rowcolor{stripblue}
\textbf{Estimation strategy} 
    & \textbf{Loss} 
    & \textbf{Method} 
    & \textbf{Joint penalty} \\
\midrule

\multirow{5}{2.5cm}{\raggedright Penalised likelihood}
& \multirow{5}{*}{%
    $\displaystyle\min_{\{\Omega^{(k)}\}} \sum_{k=1}^{K} n_k \Big[ 
    \operatorname{tr}\!\big(\hat\Sigma^{(k)} \Omega^{(k)}\big) - \log\det\big(\Omega^{(k)}\big) \Big]$}
& Fused Graphical Lasso (FGL) 
& $\displaystyle
  \lambda_1 \sum_{k=1}^K \sum_{i\neq j} \left|\Omega^{(k)}_{ij}\right| + 
  \lambda_2 \sum_{k < k'} \sum_{i,j} \left|\Omega^{(k)}_{ij} - 
  \Omega^{(k')}_{ij}\right|$ \\[6pt]

& & Group Graphical Lasso (GGL) 
& $\displaystyle
  \lambda_1 \sum_{k=1}^K \sum_{i\neq j} \left|\Omega^{(k)}_{ij}\right| + 
  \lambda_2 \sum_{i \neq j} \sqrt{\sum_{k=1}^K 
  \left(\Omega^{(k)}_{ij}\right)^2}$ \\[6pt]

& & Perturbed-Node JGL (PNJGL) 
& $\displaystyle
  \lambda_1 \sum_{k=1}^K \sum_{i\neq j} \left|\Omega^{(k)}_{ij}\right| + 
  \lambda_2 \sum_{k < k'} \left\|V^{(kk')}\right\|_2$,
  \newline s.t.\ $\Omega^{(k)} - \Omega^{(k')} = 
  V^{(kk')} + V^{(kk')^\top}$ \\[6pt]

& & Independent GL 
& $\displaystyle
  \lambda_1 \sum_{k=1}^K \sum_{i\neq j} 
  \left|\Omega^{(k)}_{ij}\right|$ \\[6pt]

& & Pooled GL 
& \multicolumn{1}{l}{%
    Single graph fitted on all observations.} \\

\midrule

\multirow{4}{2.5cm}{\raggedright Nodewise Lasso regression}
& \multirow{4}{5.2cm}{%
    $\displaystyle\min_{\{\beta_j^{(k)}\}} \sum_{k=1}^{K}\frac{1}{2n_k}
    \left\|X^{(k)}_j - X_{-j}^{(k)}\beta^{(k)}_{j}\right\|_2^2$}
& Fused Neighbourhood Selection (FNS) 
& $\displaystyle
  \lambda_1\sum_{k=1}^K\left\|\beta^{(k)}_{j}\right\|_1 + 
  \lambda_2\sum_{k < k'}\left\|\beta^{(k)}_{j} - 
  \beta^{(k')}_{j}\right\|_1$ \\[6pt]

& & Group Neighbourhood Selection (GNS) 
& $\displaystyle
  \lambda_1\sum_{k=1}^K\left\|\beta^{(k)}_{j}\right\|_1 + 
  \lambda_2\sum_{i \neq j}\sqrt{\sum_{k=1}^K 
  \left(\beta^{(k)}_{ij}\right)^2}$ \\[6pt]

& & Independent NS 
& $\displaystyle
  \lambda_1\sum_{k=1}^K\left\|\beta^{(k)}_{j}\right\|_1$ \\[6pt]

& & Pooled NS 
& \multicolumn{1}{l}{%
    Single graph fitted on all observations} \\

\midrule

\multirow{1}{2.5cm}{\raggedright Proposed method}
& \multirow{1}{5.2cm}{%
    $\displaystyle\min_{\boldsymbol{\theta_j},\{\boldsymbol{\Delta_j^{(k)}}\}} 
    \sum_{k=1}^{K}\frac{1}{2N}
    \left\|X^{(k)}_j - X_{-j}^{(k)}(\boldsymbol{\theta_j} + 
    \boldsymbol{\Delta_j^{(k)}})\right\|_2^2$}
& Data Shared Neighbourhood Selection (DSNS) 
& $\displaystyle
  \lambda_1\sum_{k=1}^K\left\|\boldsymbol{\Delta_j^{(k)}}\right\|_1 + 
  \lambda_2\left\|\boldsymbol{\theta_j}\right\|_1$ \\

\bottomrule
\end{tabularx}
\vspace{6pt}
\caption{Summary of all methods considered in the simulation study. All joint methods involve two regularisation parameters. For the fused and group estimators, $\lambda_1$ controls within-condition sparsity, and $\lambda_2$ controls the degree of similarity enforced across conditions with $\lambda_2 = 0$ corresponding to the independent variants. The correspondence differs for DSNS, $\lambda_1$ penalizes the condition-specific deviations and $\lambda_2$ penalises the shared component. Pooled estimators fit a single network on the combined datasets, ignoring condition membership.}
\label{table1}
\end{sidewaystable}

\begin{figure}[h]%
\centering
\includegraphics[width=0.99\textwidth]{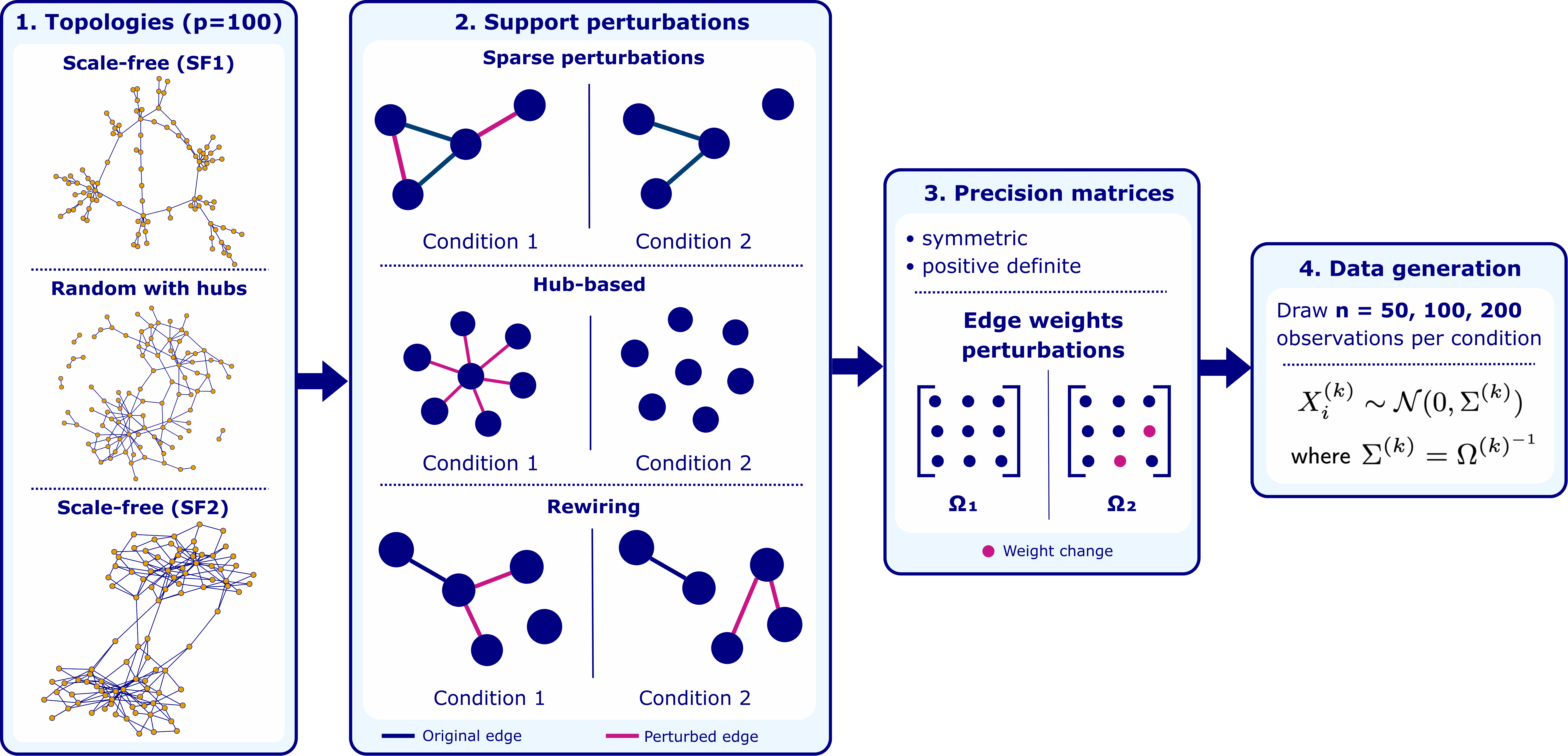}
\caption{Overview of the simulation design 1.}\label{fig_simudesign}
\figalttext[Schematic flow diagram of simulation design in four steps.]{Schematic flow diagram in four steps. Step 1 shows three network topologies for p = 100 nodes: a scale-free graph with linear preferential attachment, a random graph with added hubs, and a denser scale-free graph with preferential attachment parameter 2. Step 2 illustrates three support perturbation mechanisms between conditions 1 and 2: sparse perturbation, hub-based removal, and rewiring, with perturbed edges highlighted. Step 3 shows two precision matrices as grids of entries, differing in a small number of edge weights. Step 4 indicates data generation from a multivariate normal distribution with covariance the inverse precision matrix, at sample sizes 50, 100 and 200 per condition.}%
\end{figure}

We evaluated DSNS in two complementary simulation settings. First, we considered a two-condition setting, reflecting the case-control structure of our application. Second, to assess the behaviour of DSNS in multi-condition problems, we considered $K=3$ and $K=5$ related networks generated from a common base graph with sparse deviations for each condition. 

\paragraph*{Simulation design 1: two-condition setting}
\autoref{fig_simudesign} illustrates the simulation design. For the two-condition setting, simulations were generated for $p=100$ variables. To generate the network supports, three network topologies were considered to represent sparse networks with different degree heterogeneity: random network with added hubs (RH), scale-free network with linear preferential attachment (SF1) and scale-free with preferential attachment parameter 2 (SF2), yielding a denser graph.

Starting from a common graph structure, we introduced support differences between the two conditions using three representative perturbation mechanisms: (i) sparse perturbation, in which 20\% of edges differ in support and 20\% of precision matrix entries differ in value between conditions, (ii) rewiring, in which 20\% of edges are rewired and (iii) hub support perturbations, in which edges connected to a hub node are all removed in one condition and preserved in the other. Additional experiments varying the proportion of rewired or perturbed edges were considered. Details on simulation procedures are provided in Supplementary Material~\ref{app_simusetup}. For each resulting graph support, a sparse positive-definite precision matrix was generated. A fourth perturbation scenario was considered: hub value perturbations, in which the weights of the edges connected to a hub node are all modified for one condition while they maintain a common graph structure.

Each scenario was evaluated under two sample sizes, $n_k=50$ and $200$. The choice of $n_k=50$ reflects the limited sample sizes commonly encountered in omics studies, while $n_k=200$ was selected to approximate the sample size available in our proteomics application. For each condition, observations were drawn independently from a multivariate Gaussian distribution $\mathcal{N}(0, \Sigma^{(k)})$, 
where $\Sigma^{(k)} = \Omega^{(k)^{-1}}$ is the condition-specific 
covariance matrix. Performance was additionally assessed at $n_k = 100$. Each configuration was replicated 50 times.

\paragraph*{Simulation design 2: multi-condition setting}
Similarly, for the multi-condition simulations each related networks were generated from a common base graph with $p=100$ nodes for all three topologies. Sparse condition-specific perturbations affecting 20\% of the original support and 20\% of precision matrix entries was distributed among the graphs to reflect settings in which several biological states share most molecular associations but each state may contain a limited number of specific alterations. Precision matrices and observations were generated as in the first simulation, with $n_k=100$ observations per condition and 50 replicates per configuration.

\paragraph*{Edge recovery performance}
Edge recovery performance was evaluated using the area under the precision-recall curve (AUPR) separately for: overall graph recovery, differential support recovery, and differential edge-weight recovery. Overall graph recovery was computed separately for each condition-specific graph and averaged over the $K$ conditions. Details are provided in Supplementary Material~\ref{app_simumetrics}.

\paragraph*{Calibration performance}
To assess the calibration of DSNS, we compared the F1 score obtained at the calibrated parameter values against the oracle F1 score corresponding to the maximum achievable performance over the regularisation grid. Calibration performance was evaluated separately for: overall graph recovery, differential support and differential precision recovery. For each simulation replicate and scenario, we computed the relative difference between oracle and calibrated F1 scores:
$$\delta = \frac{F1_{\text{oracle}} - F1_{\text{calibrated}}}{F1_{\text{oracle}}}
$$ and reported the median across simulation replicates and scenarios.

\subsection{Data overview}

We revisited a previous cohort study investigating circulating inflammatory proteins in plasma samples and their association with future lung cancer risk \citep{dagnino2021prospective}. The data came from two prospective cohorts, EPIC-Italy and the Norwegian Women and Cancer Study (NOWAC), comparing participants who subsequently developed lung cancer during follow-up with matched controls who remained lung cancer free. While the original analysis focused on identifying individual proteins associated with disease risk using penalised logistic regression, our objective was to characterise differences in the conditional dependence structure of inflammatory proteins between future lung cancer cases and matched controls.

Data included 648 prediagnostic blood samples, comprising 325 future lung cancer cases and 323 controls matched on age, sex, year of recruitment, season of blood collection and study centre. A panel of 92 inflammatory proteins was measured using an Olink assay. Following the preprocessing and quality-control procedures described in \citet{dagnino2021prospective}, 71 proteins were retained for analysis. Protein concentrations were expressed as normalised protein expression values and log-transformed before analysis. Consistent with the original study, the present analysis was restricted to women, resulting in 384 participants, including 191 future cases and 193 controls. Protein measurements were adjusted for age, body mass index and cumulative smoking exposure (pack-years).

We applied DSNS to the full set of 71 inflammatory proteins to jointly estimate the conditional independence networks of future lung cancer cases and controls. The method was calibrated using the proposed extension of StARS over a grid of 400 pairs of regularisation parameters with the instability threshold $\beta=0.05$. We explored differences between conditions in terms of both differential support, corresponding to edges present in only one of the two networks, and differential precision, corresponding to changes in conditional dependency strength. To contextualise the inferred network differences, we also reproduced the protein selection analysis from the original study of \cite{dagnino2021prospective} using Sharp stability selection \citep{bodinier2023automated} with lasso logistic regression, with future lung cancer status as the outcome. 

\section{Results}
\subsection{Simulated data}
\subsubsection{Simulation design 1: two-condition setting}
\paragraph*{Recovery of graph structures for each condition}

\begin{figure}[h]%
\centering
\includegraphics[width=0.99\textwidth]{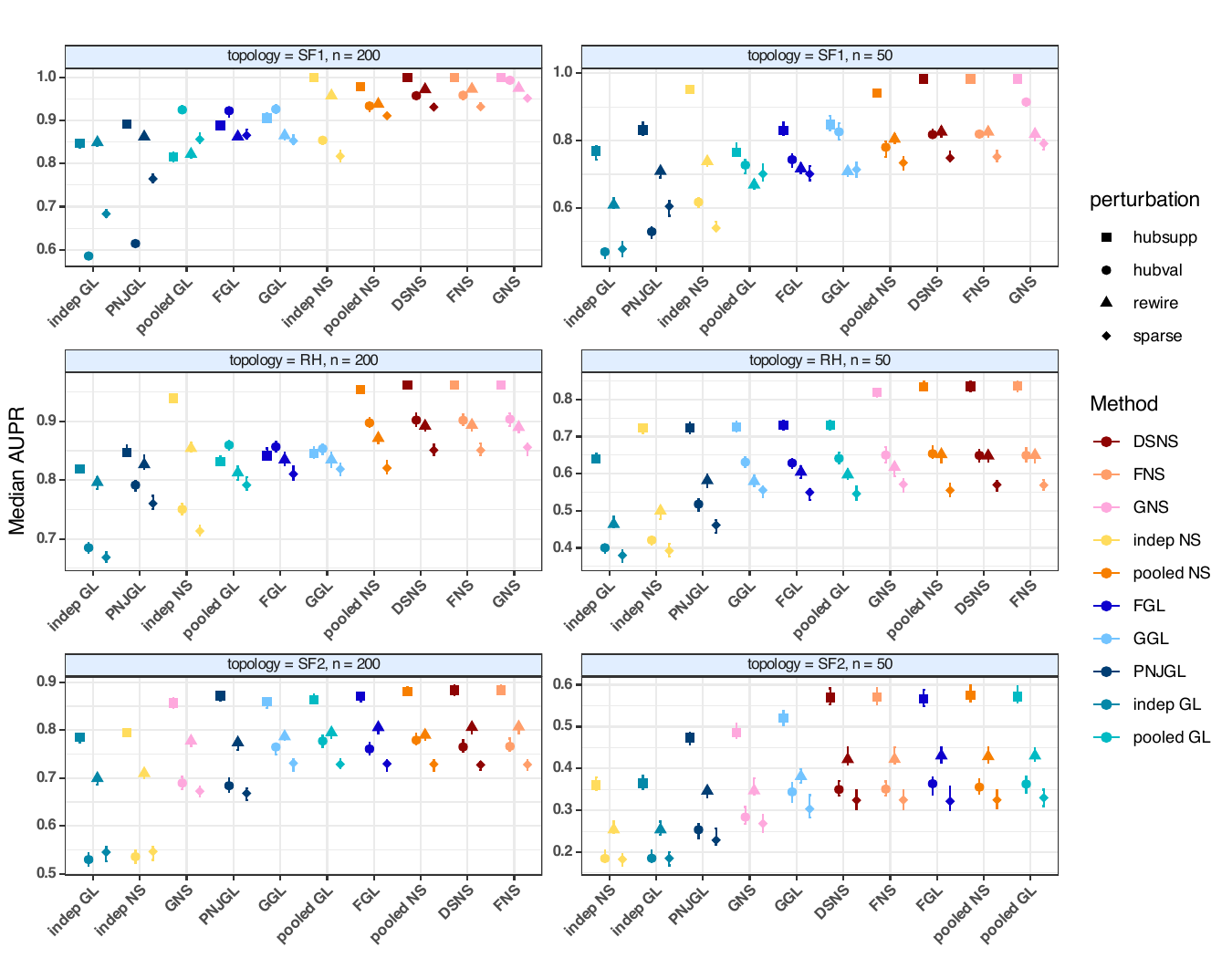}
\caption{Median area under the precision-recall curve for \textbf{graphs support recovery} across graph topologies, sample sizes and four perturbation types: 
sparse perturbations (\textit{sparse}), rewiring of edges (\textit{rewire}) and 
removal of edges around hubs (\textit{hubsupp}). Points represent median AUPR values and error bars indicate the interquartile range across simulation replicates. Methods are 
ordered by increasing median AUPR within each panel.}\label{fig_aupr_graph}
\figalttext[Two plots side by side, for n = 200 and n = 50, showing median AUPR for ten methods with interquartile range error bars.]{Two plots side by side, for n = 200 and n = 50, showing median AUPR for ten methods with interquartile range error bars. Methods are ordered by increasing median AUPR within each panel.}%
\end{figure}

The AUPR for recovering the graph structures $\mathcal{E}^{(k)}$ of each condition for all topologies and perturbations is displayed in \autoref{fig_aupr_graph}. Joint estimation consistently outperforms independent network estimations across all settings, with gains of approximately 0.15-0.20 AUPR. This confirms that borrowing information across conditions substantially improves overall graph recovery when the true networks share a common structure. DSNS preserves these gains while achieving performance comparable to, and often exceeding, that of existing joint estimation methods. Overall, DSNS consistently ranks among the top performing methods with FNS and GNS across most settings. Its performance is nearly identical to that of FNS and remains competitive with pooled estimators, indicating that the decomposition into shared and condition-specific components does not compromise overall graph recovery.

For all methods, network topology has a strong impact on recovery performance. Denser networks are more difficult to recover than sparse networks, with the SF2 topology yielding the lowest AUPR across all methods with values below 0.88 for $n_k=200$ and below 0.57 for $n_k=50$. In contrast, recovery is more accurate for the SF1 topology with AUPR above 0.8 for $n_k=200$ and above 0.7 for $n_k=50$ for all joint estimators, except PNJGL. As expected, increasing the sample size substantially improves performance across all methods and topologies.

Similar conclusions are observed for $n_k=100$, with the corresponding results reported in Supplementary \autoref{suppfig_aupr_graph}--\autoref{suppfig_aupr_valdiff}.

\paragraph*{Differential support recovery} 

\begin{figure}[h]%
\centering
\includegraphics[width=0.99\textwidth]{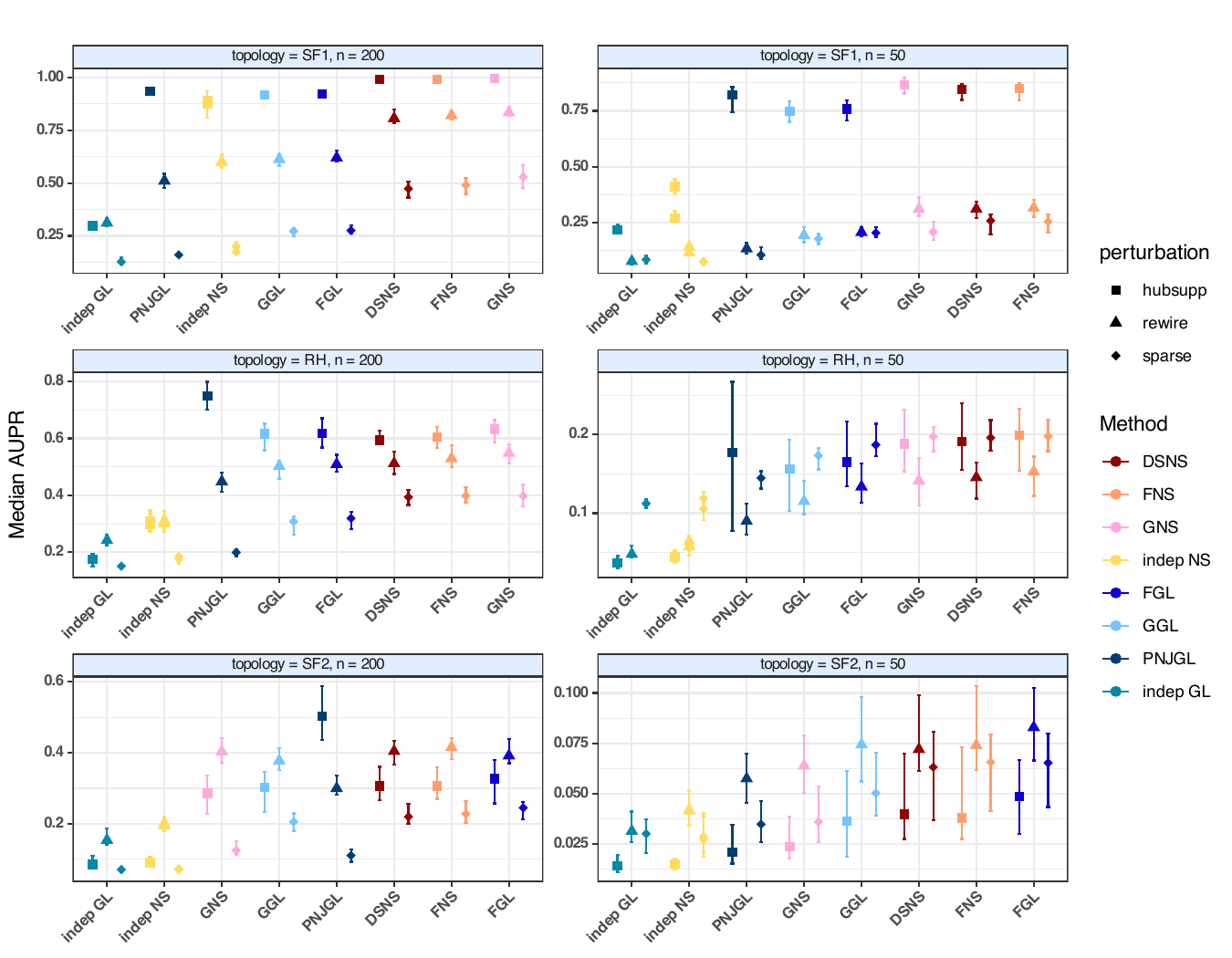}
\caption{Median area under the precision-recall curve for \textbf{differential support recovery} across graph topologies, sample sizes and four perturbation types: 
sparse perturbations (\textit{sparse}), rewiring of edges (\textit{rewire}) and 
removal of edges around hubs (\textit{hubsupp}). Points represent median AUPR values and error bars indicate the interquartile range across simulation replicates. Methods are 
ordered by increasing median AUPR within each panel.}\label{fig_aupr_diff_supp}
\figalttext[Grid of six plots: three network topologies by two sample sizes.]{Grid of six plots: three network topologies (SF1, RH, SF2) by two sample sizes (n = 200, n = 50), showing median AUPR for eight methods with interquartile range error bars. Point shape distinguishes three perturbation types: hub support removal, rewiring and sparse perturbation.}%
\end{figure}

\autoref{fig_aupr_diff_supp} presents the AUPR for recovering the differential support $\mathcal{E}_{\text{suppdiff}}$ in scenarios involving structural differences between conditions. Across all perturbations types, all joint methods outperformed independent estimations, confirming the benefit of joint estimation for differential support identification in addition to the recovery of the individual networks. 

However, recovery is strongly affected by network topology and the type of perturbation. Dense networks remain challenging for all methods, particularly in high dimensional settings. For the SF2 topology, AUPR values remain below 0.52 for $n_k=200$ and below 0.10 for $n_k=50$, indicating that reliable differential support recovery becomes difficult when the underlying graphs are highly connected.

The perturbation type also had a strong impact on performance. Differential structures concentrated around hubs are substantially easier to detect than differences arising from rewiring or sparse edge modifications for the SF1 and RH topologies. PNJGL, which explicitly encourages node differential structure, achieves the highest performance in the hub support scenario. In contrast, its advantage disappears in the rewiring and sparse perturbation settings.

In the rewiring and sparse perturbation scenarios, DSNS consistently ranks among the top three best performing methods, with performance close to or equal to FNS and GNS for the SF1 and RH topology, and performances similar to FNS and FGL in the SF2 topology. These scenarios correspond most closely to the modelling assumption underlying DSNS: networks are largely shared across conditions, with sparse and distributed deviations from a common structure. The strong performance of DSNS in these settings suggests that explicitly modelling perturbations around a shared component is well suited to detecting distributed network perturbations. Nevertheless, differential support recovery remains substantially more challenging than overall graph recovery, with AUPR values rarely exceeding 0.5.

\paragraph*{Differential edge weights recovery}

\begin{figure}[h]%
\centering
\includegraphics[width=0.99\textwidth]{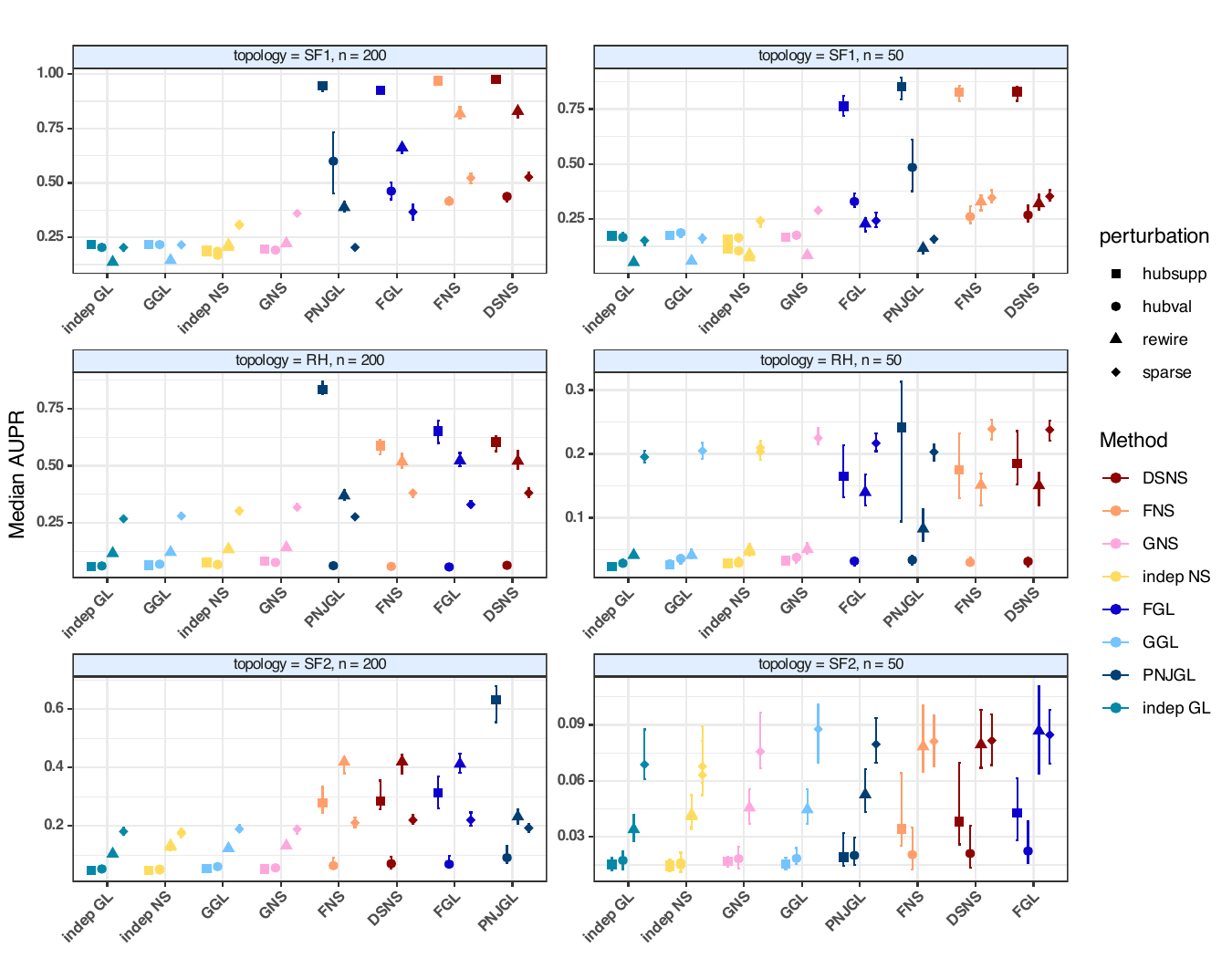}
\caption{Median area under the precision-recall curve for \textbf{differential precision edge recovery} across graph topologies, sample sizes and four perturbation types: 
sparse perturbations (\textit{sparse}), rewiring of edges (\textit{rewire}), 
removal of edges around hubs (\textit{hubsupp})  and differences of edge values connecting a hub (\textit{hubval}). Points represent median AUPR values and error bars indicate the interquartile range across simulation replicates. Methods are 
ordered by increasing median AUPR within each panel.}\label{fig_aupr_diff_prec}
\figalttext[Grid of six plots: three network topologies by two sample sizes, showing median AUPR for eight methods with interquartile range error bars.]{Grid of six plots: three network topologies by two sample sizes, showing median AUPR for eight methods with interquartile range error bars. Point shape distinguishes four perturbation types: hub support removal, hub value change, rewiring and sparse perturbation.}%
\end{figure}

The AUPR for recovering the differential precision edges $\mathcal{E}_{\text{precdiff}}$ is shown in \autoref{fig_aupr_diff_prec}. Similar patterns are observed to those seen for differential support recovery. In particular, denser graph structures lead to much lower performance, with the impact becoming more pronounced in high dimensional settings.

The performance of the methods depends strongly on the type of perturbation. In the hub support scenario, PNJGL achieves the highest AUPR across most settings, consistent with its joint penalty. In contrast, when differences arise through rewiring or sparse perturbations, DSNS consistently achieves either the highest or second highest AUPR across topologies and sample sizes, with performance comparable to FNS and FGL. These findings indicate that DSNS is particularly effective at identifying differential edge weights generated by sparse network reorganisation.

The hub value scenario, where conditions share the same support but differ only in edge weights around hub nodes, is the most challenging perturbation type for all methods. In the RH and SF2 topologies, all methods achieve AUPR values below 0.10 regardless of sample size, demonstrating the difficulty of detecting edge weight changes in the absence of structural differences. Performance improves in the simpler SF1 topology, where PNJGL and FGL achieve the highest AUPR values.

The same conclusions are observed across a broader range of rewiring levels, with the proportion of rewired edges varying from 5\% to 40\% (Supplementary \autoref{suppfig_aupr_rewire_avg}-\ref{suppfig_aupr_rewire_valdiff}).

\subsubsection{Simulation design 2: multi-condition setting}
\begin{figure}[h]%
\centering
\includegraphics[width=0.99\textwidth]{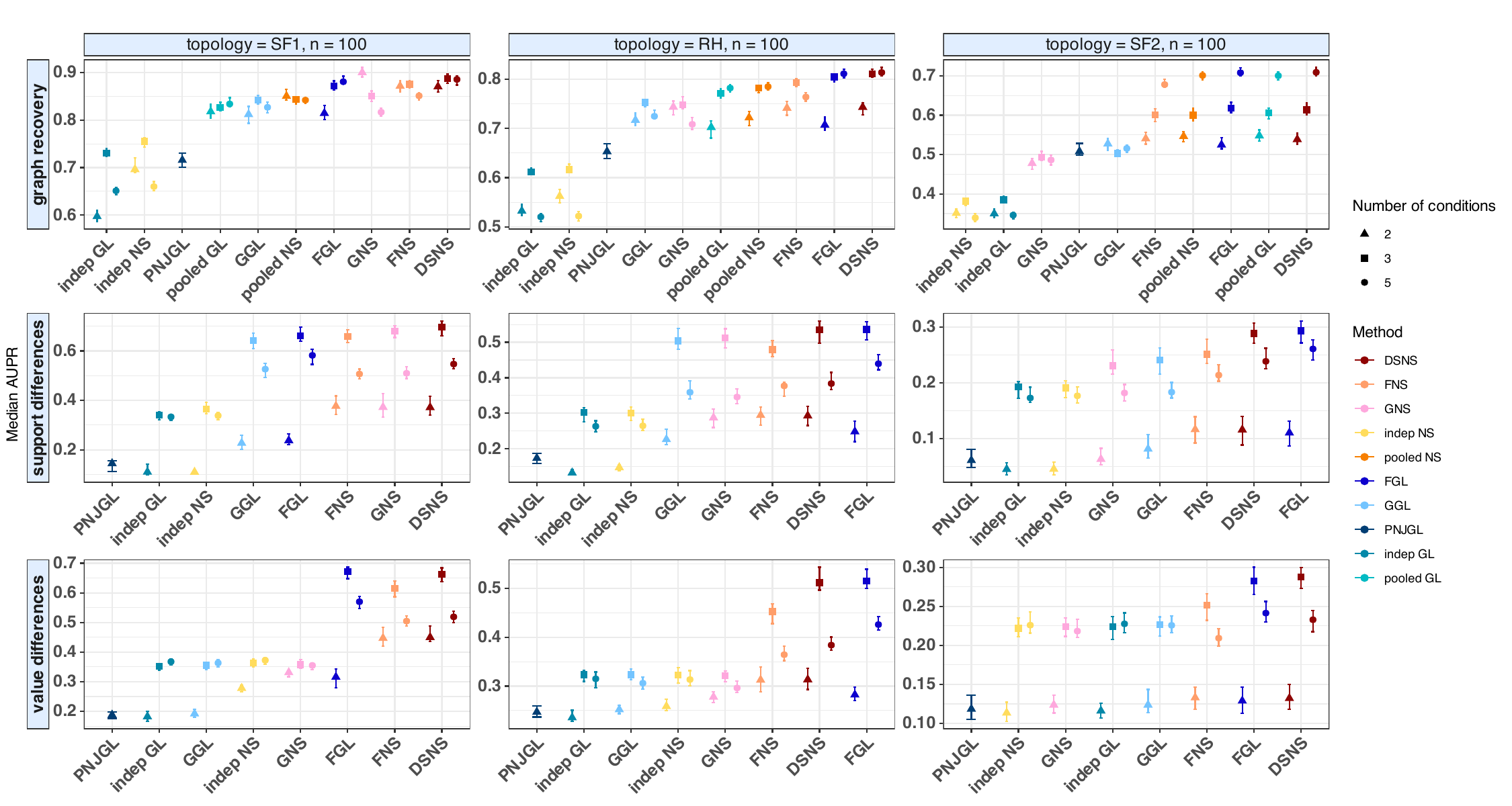}
\caption{Median area under the precision-recall curve for graphs, differential support and differential precision edge recovery across different number of conditions. Points represent median AUPR values and error bars indicate the interquartile range across simulation replicates. Methods are 
ordered by increasing median AUPR within each panel.}\label{fig_aupr_diffcond}
\figalttext[Grid of nine plots: three network topologies by three differential edge sets recovery.]{Grid of nine plots: three network topologies by three differential edge sets recovery, showing median AUPR for ten methods with interquartile range error bars. Point shape distinguishes three number of conditions: 2, 3 and 5.}
\end{figure}

\autoref{fig_aupr_diffcond} presents the AUPR for overall graph recovery, differential support recovery and differential precision recovery in the multi-condition setting with $K=2$, $3$ and $5$. Across all network topologies, DSNS consistently achieved the best overall graph recovery performance. For differential support and differential precision recovery, DSNS was either the best-performing method or ranked second after FGL in a subset of scenarios.

Compared with FNS, whose performance was comparable to DSNS in the two-condition setting, DSNS showed a clearer advantage as the number of conditions increased. This improvement is consistent with the explicit decomposition of nodewise coefficients into a shared component and sparse condition-specific deviations, rather than encouraging similarity through pairwise fusion penalties.

\subsubsection{Calibration performance}
The median relative difference between oracle and calibrated F1 scores across simulation replicates and scenarios for DSNS is reported in \autoref{tab_calib}. Smaller values indicate that the calibration method selects a parameter pair closer to the oracle choice.
StARS calibration with $\beta=0.05$ gave the lowest average gap to the oracle F1 score for graph,  differential support and differential precision recovery.
The difference with other calibration methods was especially pronounced for overall graph recovery, where AIC, BIC and eBIC selected substantially less favourable parameter values on average. Although eBIC performed competitively for differential support and differential edge-weight recovery, StARS with $\beta=0.05$ provided the most reliable calibration overall.

\begin{table}[h]
\centering
\setlength{\aboverulesep}{0pt}
\setlength{\belowrulesep}{0pt} 
\setlength{\extrarowheight}{0.4ex} 
\renewcommand{\arraystretch}{1.1}
\begin{tabular}{lccc}
\toprule
\rowcolor{stripblue}
\textbf{Calibration method} 
    & \textbf{Graph recovery} 
    & \textbf{Differential support} 
    & \textbf{Differential edge weights} \\
\midrule
StARS ($\beta = 0.05$) & $\mathbf{0.092}$ & $\mathbf{0.265}$ & $\mathbf{0.237}$ \\
StARS ($\beta = 0.10$) & $0.288$ & $0.632$ & $0.552$ \\
eBIC ($\gamma = 0.5$)  & $0.314 $& $0.335$ & $0.301$ \\
BIC                    & $0.314$ & $0.620$ & $0.575$ \\
AIC                    & $0.805$ & $0.931$ & $0.897$ \\
\bottomrule
\end{tabular}
\vspace{5pt}
\caption{Calibration performance for DSNS, measured by the median relative gap between the calibrated F1 score and the oracle F1,
averaged across scenarios and simulations. Results are reported separately for graph recovery, differential support recovery, and differential edge weight recovery. Values closer to zero indicate better calibration. The smallest value for each edge set is highlighted in bold.}
\label{tab_calib}
\end{table}

\subsubsection{Computational performance}

\begin{table}[h]
\centering
\setlength{\aboverulesep}{0pt}
\setlength{\belowrulesep}{0pt}
\setlength{\extrarowheight}{0.4ex}
\begin{tabular}{lcccccc}
\toprule
\rowcolor{stripblue}
\textbf{Method} & \multicolumn{3}{c}{\textbf{Median Time (ms)}} & \multicolumn{3}{c}{\textbf{Total Memory Allocated (MB)}} \\
\rowcolor{stripblue} 
 & $p=100$ & $p=300$ & $p=500$ & $p=100$ & $p=300$ & $p=500$ \\
\midrule
DSNS (Ours)& 122.12          & \textbf{760.69}            & \textbf{2 099.62}           & 224.32          & 1 961.36          & 5 418.505           \\
GNS        & \textbf{96.17}  & 781.36   & 2 230.50  & \textbf{168.64} & \textbf{1 472.67} & \textbf{4 049.35}  \\
FNS      &  5 983.34 & 145 933.21 & 668 316.18 & 251.55 & 5 093.33 & 22 118.46 \\
GGL        & 345.21          & 8 143.58           & 29 486.09          & 389.03          & 3 321.75          & 11 789.31          \\
FGL        & 201.11          & 9 278.09           & 28 578.99          & 283.99          & 6 088.28          & 13 677.46          \\
PNJGL      & 588.52          & 13 023.36          & 47 403.15          & 715.32          & 5 955.95          & 18 185.91          \\
\bottomrule
\end{tabular}
\vspace{5pt}
\caption{Median execution time and memory allocation for each method across 
number of variables $p \in \{100, 300, 500\}$ for $K=2$ and $n_k=100$.}
\label{tab_compute}
\end{table}

Computational time and memory allocation across candidate methods are reported in \autoref{tab_compute}. Among the neighbourhood selection approaches, DSNS and GNS showed similar computational increase as the network increased and remained considerably faster than FNS and graphical lasso methods whose computation time increased more substantially with $p$. DSNS had computational time comparable to GNS, and was slightly faster at $p=300$ and $p=500$, which may be explained by the simpler penalty structure. Regarding memory allocation, DSNS required more memory than GNS due to the extended design matrix whose column dimension grows as $(p-1)(K+1)$. Nevertheless, memory usage remained substantially lower than the graphical lasso methods. Overall, DSNS achieved a computational performance that makes it scalable to high-dimensional problems and advantageous over other methods.

\subsection{Application: inflammatory protein network reorganisation in future lung cancer cases}

\begin{figure}[H]
      \parbox{0.9\linewidth}{
        \centering
        \includegraphics[width=1\linewidth]{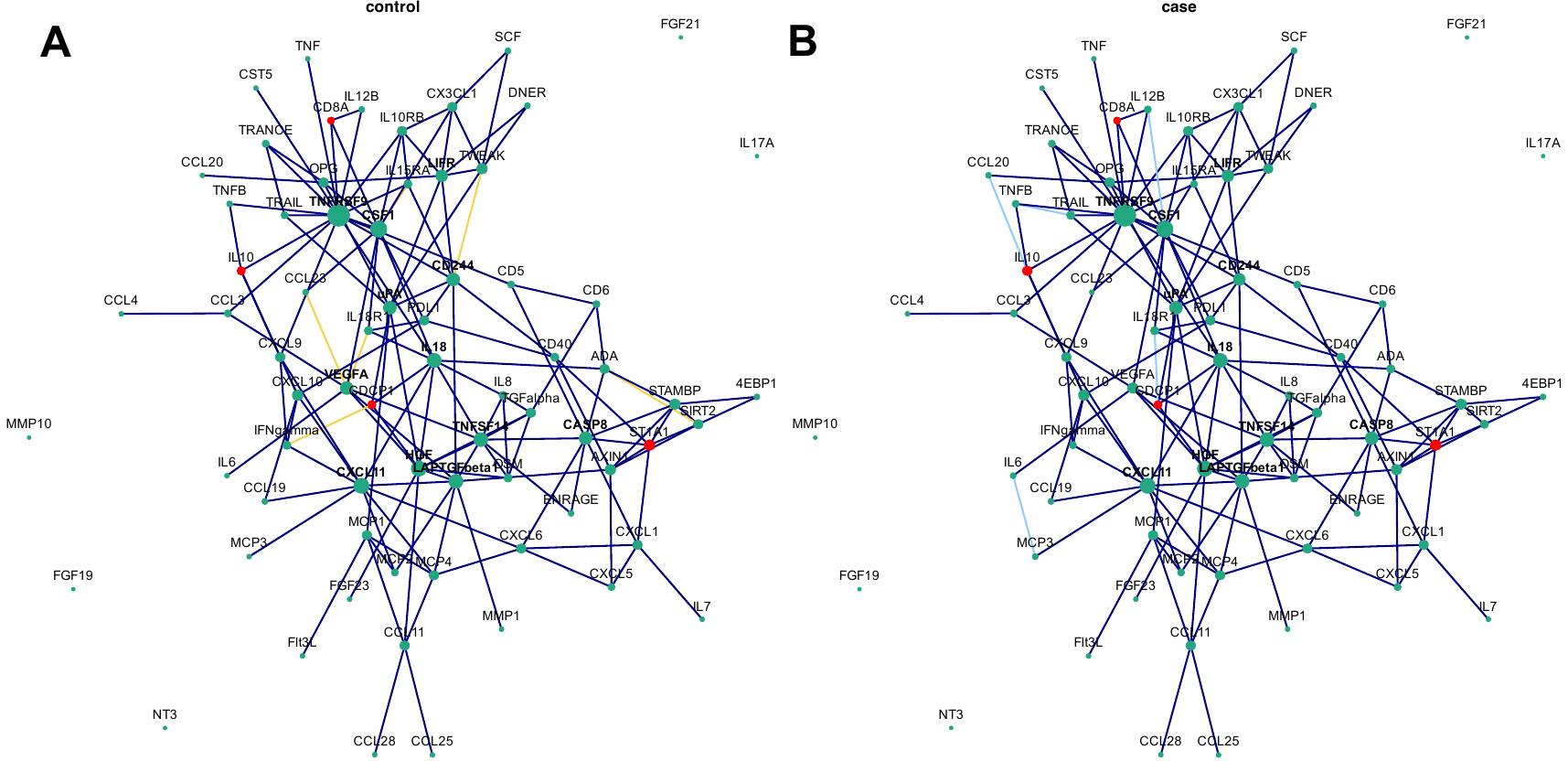}
    }
    \parbox{0.5\linewidth}{
        \centering
        \includegraphics[width=1\linewidth]{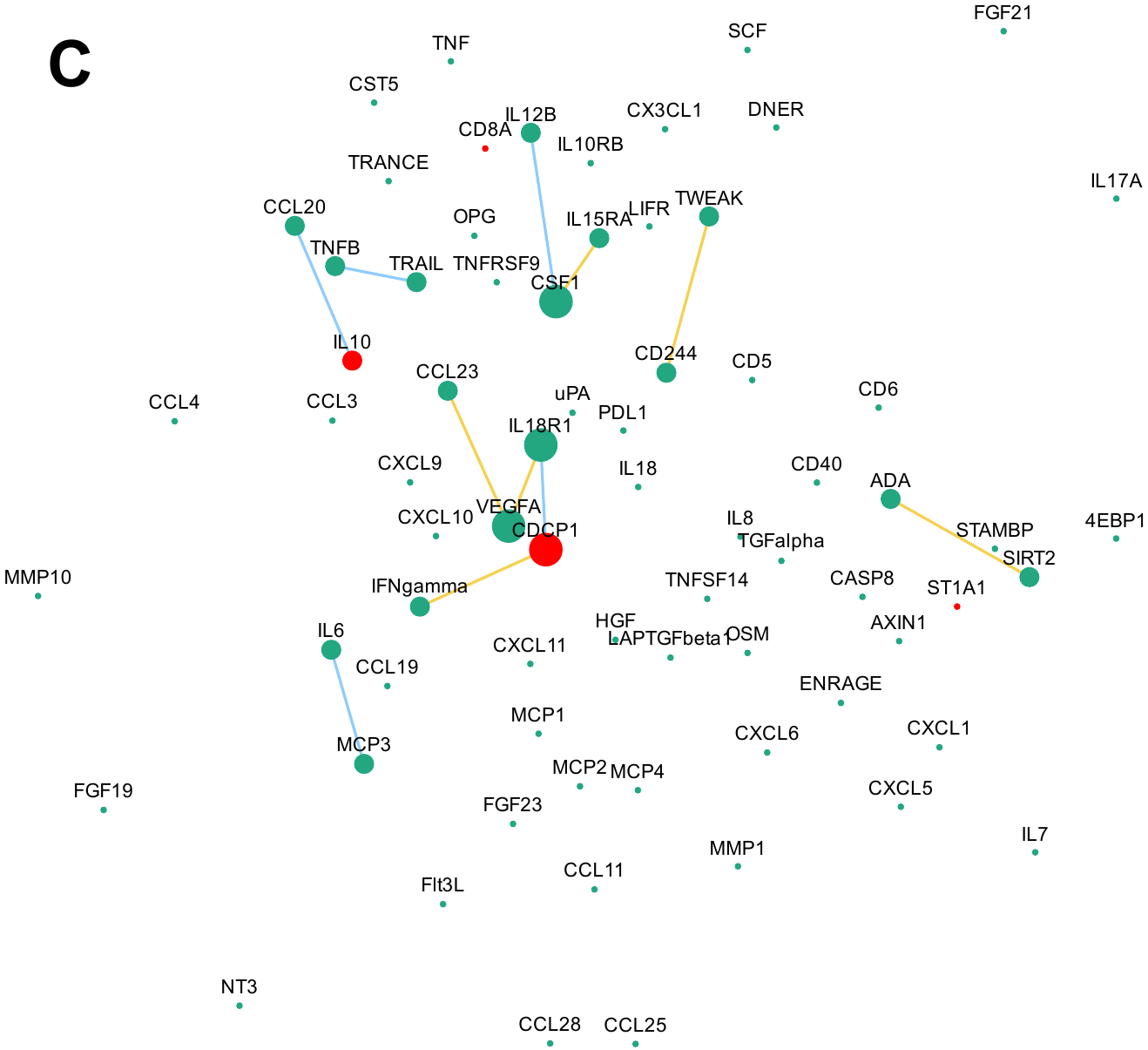}
    }
    \parbox{0.5\linewidth}{
        \centering
        \includegraphics[width=1\linewidth]{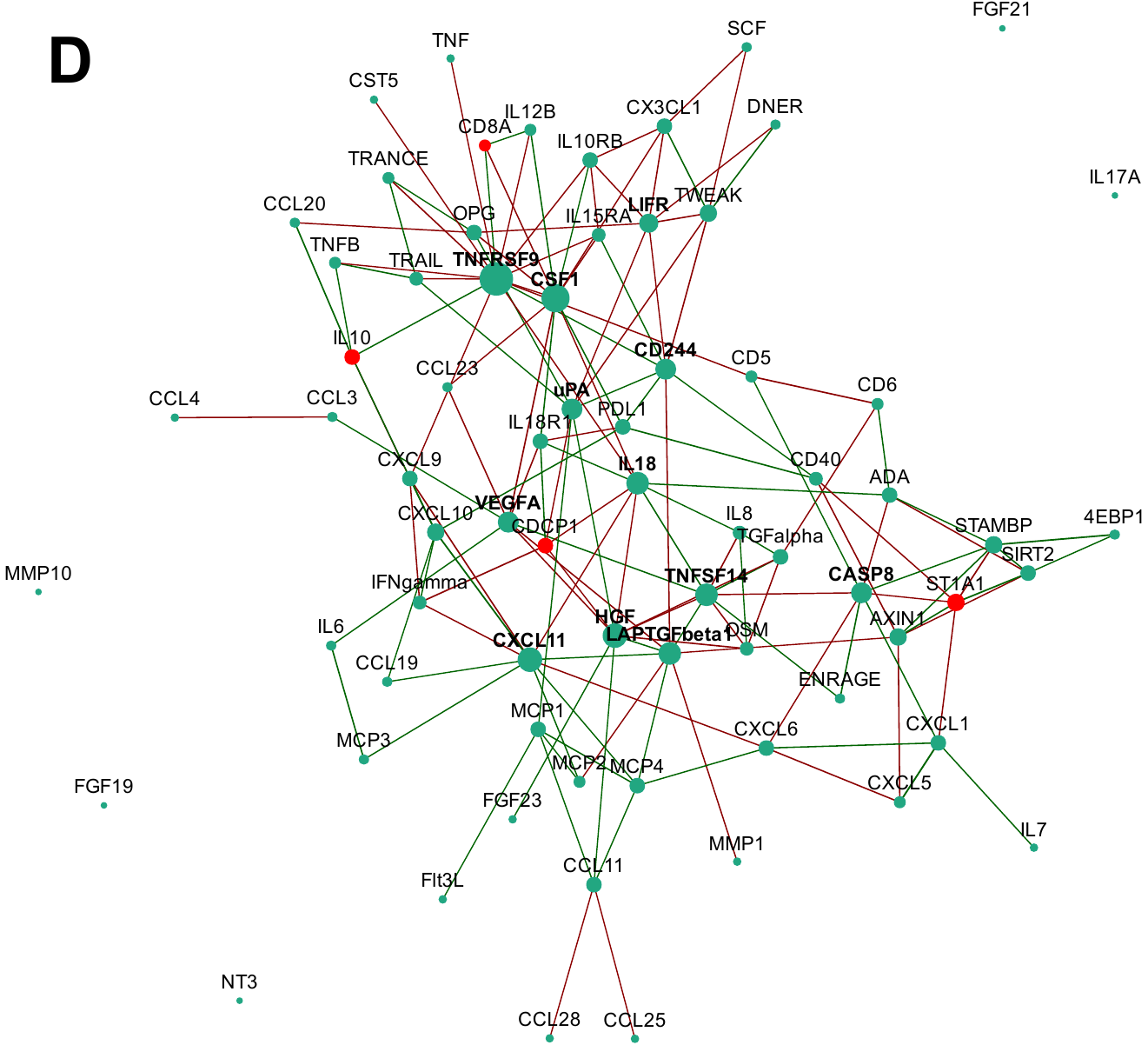}
    }
    \caption{ Estimated graph supports for (A) controls and (B) future lung cancer cases, inferred using the \textit{DSNS} method. (C) Estimated differential support and (D) differential precision matrix of the inflammatory protein network between future lung cancer cases and controls. Node size is proportional to node degree. Red nodes were identified by logistic regression as stably associated with lung cancer risk, while green nodes were not identified as associated with lung cancer risk. Bold node labels indicate hub nodes with more than 6 edges in the graph. Dark blue edges are common to both conditions, yellow edges are specific to controls, and light blue edges are specific to future lung cancer cases. In the differential precision matrix (D), edge colors indicate the sign of the difference between the precision matrices: green edges indicate a larger conditional dependency in controls than in future lung cancer cases, whereas red edges indicate a larger conditional dependency in future lung cancer cases than in controls.}
    \label{fig3}
\figalttext[described]{Two network diagrams of 71 inflammatory proteins. Left: the differential support network. Edges specific to controls are drawn in dark blue and edges specific to future cases in light blue. Right: the differential precision network, which is far denser, with most nodes connected; edge colour indicates the sign of the precision difference, green where the conditional dependency is stronger in controls and red where it is stronger in cases. In both panels node size is proportional to degree, and the four proteins selected by logistic regression as associated with lung cancer risk: IL10, CDCP1, ST1A1 and CD8A, are shown in red, all others in green.}
\end{figure}

The networks estimated by DSNS are displayed in \autoref{fig3}A-B. The control network included 146 edges and the future lung cancer cases network included 145 edges. Among these, 140 edges were shared across conditions, while 11 edges were present in one condition but absent in the other: six were specific to controls and five were specific to future cases. The Sharp calibrated lasso logistic regression selected four proteins associated with future lung cancer risk: IL10, CDCP1, ST1A1 and CD8A.

The estimated differential support network is displayed in \autoref{fig3}C. Four proteins CDCP1, CSF1, IL18R1 and VEGFA displayed the largest number of differential associations, with two each. CDCP1 was also selected in the logistic regression analysis as associated with future lung cancer risk, whereas CSF1, IL18R1 and VEGFA were not. This suggests that the network analysis may highlight proteins whose relevance is expressed through altered conditional associations rather than through marginal association with disease status. The association between CDCP1 and IFNgamma was present in controls but absent in future cases, whereas associations between CDCP1 and IL18R1 was specific to future cases. IL10 also selected in the logistic regression analysis, displayed an additional association in future cases with CCL20. These results suggest that proteins previously identified as individual risk markers may also be embedded in broader alterations of the inflammatory association structure preceding lung cancer diagnosis.

The differential precision network which contains changes in conditional dependency strength is displayed in \autoref{fig3}D. Several associations displayed substantial differences despite being present in both networks. The largest absolute association differences were observed for CXCL1-CXCL5, 4EBP1-STAMBP, CXCL11-CXCL10 and CXCL9-CXCL10. Although these proteins were not identified as disease markers, these changes in network organisation may be involved in inflammatory dysregulation. The ten largest absolute differences are reported in \autoref{tab_diff_prot}. These included CDCP1-IFNgamma and IL10-CCL20 which were also identified in the differential support, confirming that these associations are highly perturbed between lung cancer cases and controls.

\begin{table}[h]
\centering
\setlength{\aboverulesep}{0pt}
\setlength{\belowrulesep}{0pt} 
\setlength{\extrarowheight}{0.3ex}
\renewcommand{\arraystretch}{1}
\begin{tabular}{llc}
\toprule
\rowcolor{stripblue}
\textbf{Protein 1} & \textbf{Protein 2} & \textbf{Difference}\\
\midrule
CXCL1 & CXCL5 & 0.993 \\
 4EBP1 & STAMBP & 0.819 \\
 CXCL11 & CXCL10 & 0.759 \\
 CXCL9 & CXCL10 & 0.726 \\
 SIRT2 & ST1A1 & 0.718 \\
 IL10 & CCL20 & 0.557 \\
 CD6 & CD5 & -0.55 \\
 ST1A1 & STAMBP & -0.527 \\
 CDCP1 & IFNgamma & -0.517 \\
 VEGFA & CSF1 & -0.511 \\
\bottomrule
\end{tabular}
\vspace{5pt}
\caption{Top 10 protein pairs with the largest absolute differences in estimated 
precision matrix entries between controls and cases. Positive values indicate stronger conditional dependencies in controls, negative values indicate stronger conditional dependencies in cases.}
\label{tab_diff_prot}
\end{table}

\section{Discussion}

We proposed Data Shared Neighbourhood Selection, a joint network inference framework that combines the Data Shared Lasso decomposition with neighbourhood selection to estimate related Gaussian graphical models. By representing each nodewise regression coefficient as the sum of a shared component and sparse condition-specific deviations, DSNS directly leverages the assumption that biological conditions share most molecular associations, with only a limited number of perturbations between conditions. The resulting extended lasso formulation makes the method computationally efficient, scalable to high-dimensional settings and straightforward to implement using existing optimisation algorithms. We also adapted StARS to the two-parameter joint estimation setting and showed in simulations that stability-based calibration provides more reliable graph selection than standard information criteria.

In the lung cancer proteomic application, DSNS identified differential inflammatory protein associations involving CDCP1 and IL10, two proteins reported as risk markers for lung carcinogenesis. This suggests that these markers may not act as isolated molecular signals, but may be involved in a broader reorganisation of the inflammatory association network. The analysis also highlighted CSF1, IL18R1 and VEGFA, which were not selected as individual risk markers but appeared among the most differentially connected proteins. These findings illustrate the complementary information provided by network analyses, where disease alterations may be reflected by changes in conditional associations even when individual marker effects are modest.

In the two-condition setting, DSNS achieved graph recovery performance comparable to that of FNS across a wide range of network topologies, perturbation mechanisms and sample sizes. This similarity is expected, as demonstrated in \citep{ollier2017regression}, the two objectives share the same fusion term and differ in how the sparsity penalty is applied. However, the two methods are not equivalent when $K>2$. DSNS achieved the best overall graph recovery across all three topologies and was either the best-performing method or second to FGL for differential support and differential precision recovery while FNS had lower performances than DSNS across scenarios. Indeed, unlike FNS which leverages a pairwise fusion penalty, DSNS directly provides an explicit decomposition of each condition's coefficients into shared and deviation components, providing a representation that is directly aligned with the objective of distinguishing preserved from altered molecular associations. Additionally, DSNS was particularly competitive when network differences were sparse and distributed across the graph, whereas methods designed for localised node perturbations, such as PNJGL, performed best when differences were concentrated around hubs. Finally, DSNS offered a clear computational advantage over other graphical-lasso-based joint methods.

Several limitations should be noted. Although the initial DSNS estimation decomposes nodewise coefficients into shared and condition-specific components, the constrained graphical lasso refitting step is performed independently within each condition. Differential precision matrices are therefore obtained through a posteriori comparison of refitted precision estimates and do not preserve the original shared-deviation decomposition. Additionally, since the graph support is selected and refitted using the same data, the resulting precision matrices may suffer from post-selection bias. Developing a joint unbiased refitting procedure that estimates shared and condition-specific precision matrices directly would be a valuable extension. Additionally, DSNS, like most joint graphical model estimators, does not currently provide formal uncertainty quantification for selected shared or differential edges. Incorporating confidence measures or statistical tests would improve the reproducibility and interpretability of inferred biological networks.

An extension of DSNS would be to decompose not only the nodewise regression coefficients, but also the condition-specific intercepts. In this work, variables were centered within each condition, so DSNS focuses on differences in conditional association structure rather than differences in protein abundance. Modelling the intercepts jointly would allow the identification of proteins whose conditional mean differs between biological states after adjustment for the remaining proteins, alongside protein pairs whose conditional associations differ. Such an extension would provide a unified framework for combining multivariable marker selection and joint network inference.

\section{Conflicts of interest}
MC-H holds shares of the O-SMOSE company. Consulting activities of the company are independent of the present work. The other authors declare that they have no competing interests.

\section{Funding}
BF, RC-Z, VV, JC and MC-H acknowledge support from the European Union DISCERN project (grant agreement 101096888). MC-H was also supported by the Horizon Europe Innovation programme STAGE project (grant agreement 101137146) and accompanying UKRI grant number 10109957. Views and opinions expressed are however those of the authors only and do not necessarily reflect those of the European Union. Neither the European Union nor the granting authority can be held responsible for them.

Where authors are identified as personnel of the International Agency for Research on Cancer/World Health Organization, the authors alone are responsible for the views expressed in this article and they do not necessarily represent the decisions, policy, or views of the International Agency for Research on Cancer/World Health Organization.

\section{Data availability}
The scripts and simulated datasets used to generate the results are available at \url{https://forge.inrae.fr/blanche.francheterre/dsns_paper_simulations}.  The R package \textit{monique} \url{https://forge.inrae.fr/blanche.francheterre/monique} implements FNS, GNS and DSNS.
The data in EPIC that support the findings of this study are available from the corresponding author upon reasonable request. The NOWAC data cannot be shared publicly because of local and national ethical and security policy. Data access for researchers will
be conditional on adherence to both the data access procedures of the Norwegian women and Cancer Cohort and the UiT The Arctic University of Norway (Tromsø, Norway; contact via T.M. Sandanger, torkjel.sandanger@uit.no, Tonje Braaten tonje.braaten@uit.no, and Arne Bastian Wiik, arne.b.wiik@uit.no) in addition to the local
ethical committee.

\section{Author contributions statement}

BF: formal analysis, investigation, methodology, software, visualisation, writing (original draft), writing (review and editing); RC-Z, VV: methodology, review; JC: Conceptualisation, methodology, investigation, formal analysis, project administration, writing (review and editing), supervision; MC-H: Conceptualisation, methodology, investigation, formal analysis, project administration, writing (review and editing), supervision.

All authors were responsible for the decision to submit the manuscript. 

\section{Acknowledgments}

\bibliographystyle{oup-abbrvnat}
\bibliography{reference}


\begin{appendices}

\section*{Supplementary Material}
\addcontentsline{toc}{section}{Supplementary Material}
\renewcommand{\thefigure}{S\arabic{figure}}
\renewcommand{\figurename}{Supplementary Figure}
\setcounter{figure}{0} 
\renewcommand{\thesection}{S\arabic{section}}
\renewcommand{\thesubsection}{S\arabic{section}.\arabic{subsection}}

\section{Joint network inference methods}
\subsection{Joint Graphical Lasso} \label{app_jgl}
Two different joint penalties were defined by \citep{danaher2014joint}:
\begin{itemize}
  \item \textit{Fused Graphical Lasso} (FGL),  which encourages similar conditional dependency strengths across conditions:
\end{itemize}
$$\mathcal{P}(\{\Omega^{(k)}\}) =\sum_{k < k'} \sum_{i,j} \left| \Omega^{(k)}_{ij} - \Omega^{(k')}_{ij} \right|$$

\begin{itemize}
  \item \textit{Group Graphical Lasso} (GGL),  which in contrast encourages shared sparsity patterns across conditions:
\end{itemize}
$$ \mathcal{P}(\{\Omega^{(k)}\}) = \sum_{i \neq j} \sqrt{\sum_{k=1}^K \left(\Omega^{(k)}_{ij}\right)^2}$$

Additionally, we consider the perturbed-node Joint Graphical Lasso (PNJGL) whose joint penalty encourages node-wise differences, ie differences are encouraged to be around specific nodes:
\begin{align*} \mathcal{P}(\{\Omega^{(k)}\})= \sum_{k < k'} 
\left| \left| V^{(kk')}\right|\right|_2 , \text{ subject to } \Omega^{(k)} - \Omega^{(k')} = V^{(kk')} + V^{(kk')^T}
\end{align*}
where $V^{(kk')} \in \mathbb{R}^{p \times p}$.
This penalisation is particularly relevant in biological settings where network perturbations may be localized around specific hub nodes.

FGL and GGL are estimated using the R package \textit{JGL} and PNJGL with the R package {DiffGraph} \citep{zhang2018diffgraph}.

\subsection{Joint Neighbourhood Selection} \label{app_jns}

Neighbourhood Selection was introduced by \citep{meinshausen2006high} to estimate a single gaussian graphical network. For each node $j$, we solve:
\begin{equation}
\min_{\boldsymbol{\beta_j}\in \mathbb{R}^{p-1}}\frac{1}{2n}
\left\|X_j - X_{-j} \boldsymbol{\beta_j}\right\|_2^2+ \lambda\left\| \boldsymbol{\beta_j} \right\|_1 
\end{equation}
where $X_{j} \in \mathbb{R}^{n}$ is the $j$th column of $X$, $X_{-j} \in \mathbb{R}^{n \times (p-1)}$ consists of all variables of $X$ except $j$ and $\lambda$, the regularisation parameter.

The neighbourhood selection framework can be extended to the multi-condition setting but adding a joint penalty across conditions. Analogously to JGL, we consider two joint penalties:
\begin{itemize}
  \item \textit{Fused neighbourhood Selection} (FNS), which encourages similar regression coefficients across conditions:
$$\mathcal{P}(\{\beta_j^{(k)}\})=
\sum_{k < k'}  \left\|\beta^{(k)}_{j} - \beta^{(k')}_{j}\right\|_1$$
\end{itemize}
\begin{itemize}
  \item \textit{Group neighbourhood Selection} (GNS), which encourages shared support patterns across conditions:
$$ \mathcal{P}(\{\beta_j^{(k)}\}) = \sum_{i \neq j}\sqrt{\sum_{k=1}^K \left(\beta^{(k)}_{ij}\right)^2}$$
\end{itemize}

\section{Data Shared Neighbourhood Selection}
\subsection{Identifiability} \label{app_dsns_iden}
The Data Shared Lasso model decomposes the regression coefficients as
$\beta^{(k)} = \theta + \Delta^{(k)}$ for each condition $k \in \{1, \ldots, K\}$,
where $\theta$ represents a shared effect across conditions and $\Delta^{(k)}$
captures deviations from the shared effect for each condition. In the absence of regularisation, the decomposition is overparameterised. Both \citet{gross2016data} and \cite{ollier2014joint}, demonstrated that the $\ell_1$ penalty induces a preferred sparse decomposition among equivalent representations. In our model, the ratio $\tau = \frac{\lambda_1}{\lambda_2}$ controls the degree of sharing across
conditions. \citet{ollier2014joint} showed that values $\tau<\frac{1}{K}$ lead to a vanishing shared component and thus to independent neighbourhood selection, and $\tau= \frac{1}{K}$ may yield non-unique decompositions. Hence, we restrict the regularisation parameters values to values that satisfy $\tau > \frac{1}{K} \Leftrightarrow \frac{\lambda_1}{\lambda_2} > \frac{1}{K}$.

\section{Simulation study}

\subsection{Details of simulated data generation} \label{app_simusetup}

First, a binary adjacency matrix of size $p \times p$ was generated according to two biologically relevant topologies: scale-free graphs and random graph with hubs.

Construction of SF1 and SF2 topology: Scale-free graphs reflect, by construction, the presence of highly connected hub nodes and lower degree nodes, a common property observed in biological networks. Two scale-free subgraphs of $50$ nodes each were independently generated using the Barabási-Albert preferential attachment model using the R package \textit{igraph}. Two values were considered to control preferential attachement through parameter $\alpha$, 1 yielding linear attachement, widely used in simulation studies for joint network inference, and 2 yielding much denser graphs, more representative of biological networks. The two subgraphs were then connected by four randomly selected edges.

Construction of RH topology: We considered another topology with lower degree heterogeneity. Similarly, two subgraphs of 50 nodes each were constructed using a random graph model. Within each subgraph, node degrees were sampled uniformly from $\{1,2,3\}$, then two nodes per subgraph were designated as hubs and assigned a degree of 10.

The resulting adjacency matrix was duplicated to represent the two conditions. Differential structure between the two conditions was then introduced using three different methods:
(i) random edge removal: a proportion of edges varying from $5\%$ to $100\%$ was randomly removed, half from condition 1 and the other half from condition 2. Additional imbalance scenarios were considered by varying the proportion of perturbations assigned to each condition.
(ii) random edge rewiring, to simulate topological reorganisation, a proportion of edges ($5\%$ to $50\%$) were randomly reassigned in condition 2. For each selected edge $(i,j)$, the edge was removed in condition 2 and replaced by a new edge $(i,j')$ or $(i',j)$ where $i'$ and $j'$ were sampled uniformly from the remaining nodes. This scenario preserves the total number of edges while changing the network topology. (iii) hub deletion, to evaluate differential edge recovery under localized perturbations, all edges connected to the highest-degree node in the original graph were removed from condition 2. 
We also considered graphs were no structure alterations were added, in order to perturb only the edge weights between conditions: (iv) hub value perturbations and sparse edge weights perturbations.

For each condition, positive definite precision matrices $\Omega^{(k)}$ were generated from the corresponding adjacency matrices using the R package \textit{fake}. For each nonzero edge, off-diagonal entries were sampled uniformly from [0.2,0.6] with randomly assigned signs. Entries corresponding to absent edges were set to 0. Positive definiteness was ensured by diagonal dominance such that:
$$\Omega_{ii} = \sum_{j=1}^p |\Omega_{ij}| + c $$
where $c > 0$ is a parameter tuned internally.

To introduce differences in edge weights independently of support differences, a subset of shared edges across conditions was perturbed by modifying the corresponding precision matrix. Specifically, for a proportion of edges ($5\%$ to $50\%$) present in both conditions, the corresponding precision coefficients in condition 2 were modified by adding or subtracting a constant sampled uniformly from $\{0.1,0.2,0.3,0.4\}$. After introducing edge weight perturbations, we assigned a common diagonal to all precision matrices to ensure positive definiteness was not impaired. For each variable, the diagonal entry was set to the maximum, over conditions, of the sum of absolute off-diagonal entries in the corresponding row.

For each condition $k$, observations were sampled independently from the centered multivariate normal distribution with covariance $\Omega^{(k)^{-1}}$.
We considered three sample size settings: $n_k=p/2$, $n_k=p$ and $n_k=2p$, respectively. The number of observations was the same for both conditions.
For each graph configuration, 50 independent datasets were generated.

\subsection{Performance metrics} \label{app_simumetrics}
Edge recovery performance was evaluated separately for: (i) overall graph recovery, $\mathcal{E}^{(k)}$ for all $k$, (ii) differential support recovery, $\mathcal{E}_\text{suppdiff}$, and (iii) differential precision matrix recovery, $\mathcal{E}_{\text{precdiff}}$.
To evaluate the performance of each setting over the full sparsity parameters grid regardless of calibration ability, we computed the area under the precision-recall curve (AUPR) over the full parameter grid for each simulation replicate and each method. Estimates were ordered by decreasing number of selected edges, with ties broken by decreasing precision. For each edge recovery target, precision and recall were computed along this path. When several parameter values yielded the same recall, only the maximum precision was retained. Precision values were then monotonized by replacing each point with the maximum precision observed at larger recall values, resulting in a non-increasing curve. The area under the precision-recall curve (AUPR) was computed using the trapezoidal rule after adding the boundary points ((0,1)) and ((1,0)). AUPR values were first computed separately for each simulation replicate, and then summarised across replicates using the median and interquartile range.

\section{Supplementary Figures}

\begin{figure}[h]%
\centering
\includegraphics[width=0.99\textwidth]{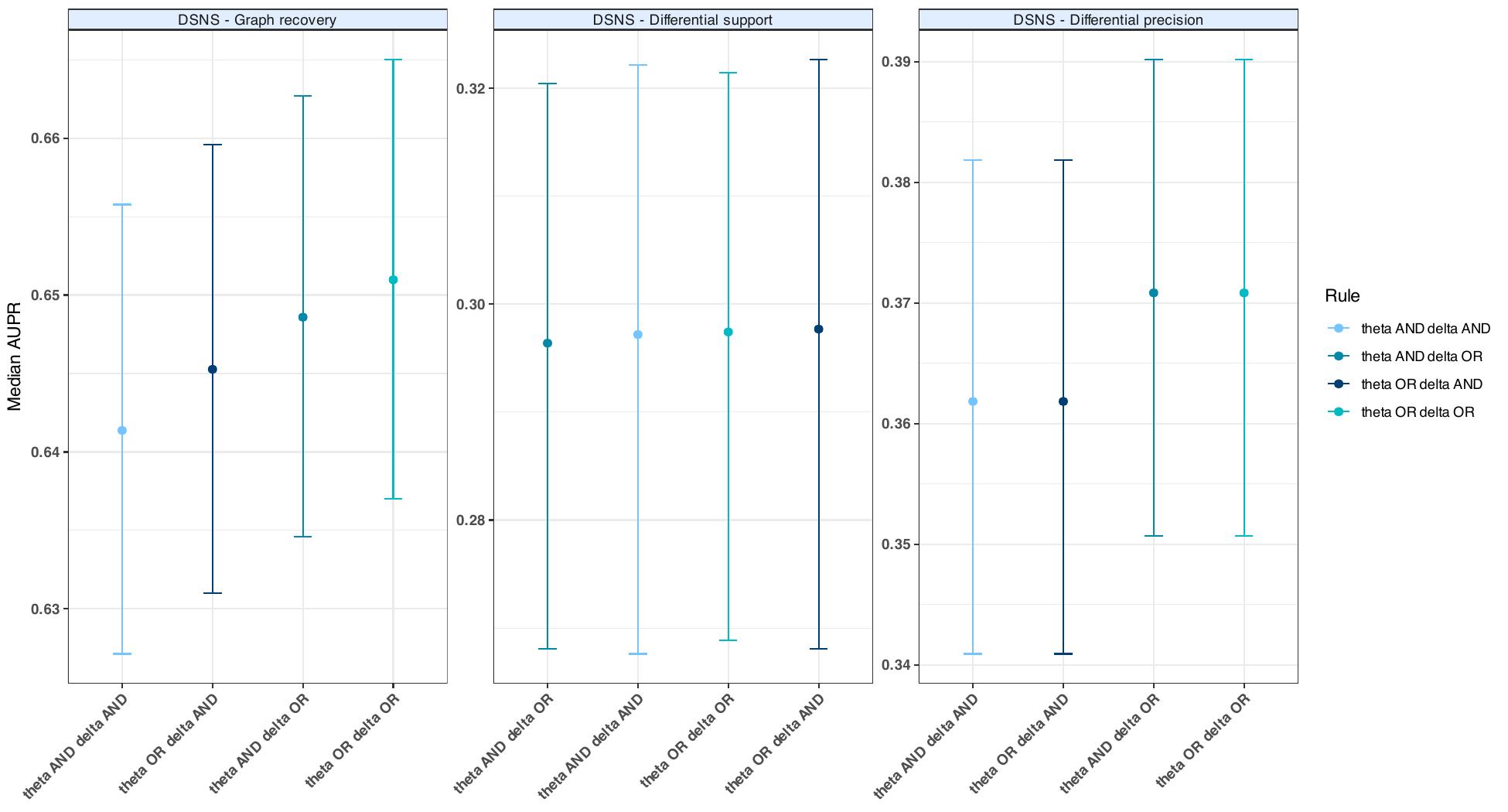}
\caption{Median area under the precision-recall curve for graphs, differential support and differential precision edge recovery, averaged over simulation scenarios, for AND, OR rule combinations for the Data Shared Neighbourhood Selection (DSNS) method.}\label{rules_dsns}
\figalttext[Three plots showing median AUPR with interquartile range error bars for graph recovery, differential support and differential precision, under four combinations of AND and OR symmetrisation rules applied to the shared and deviation components.]{Three plots showing median AUPR with interquartile range error bars for graph recovery, differential support and differential precision, under four combinations of AND and OR symmetrisation rules applied to the shared and deviation components.}%
\end{figure}

\begin{figure}[h]%
\centering
\includegraphics[width=0.99\textwidth]{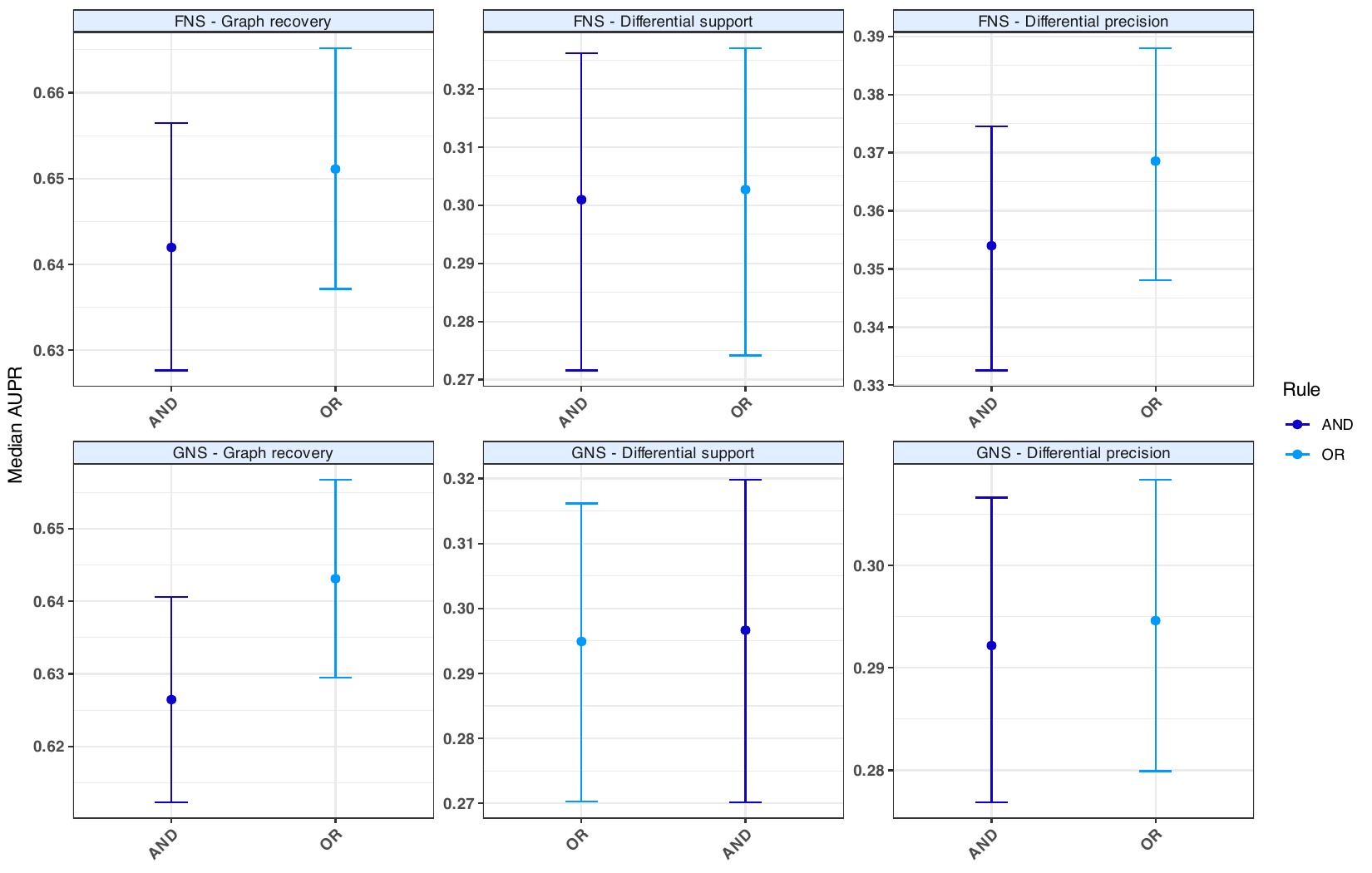}
\caption{Median area under the precision-recall curve for graphs, differential support and differential precision edge recovery, averaged over simulation scenarios, for AND, OR rules for Fused Neighbourhood Selection (FNS) and Group Neighbourhood Selection (GNS) methods.}\label{rules_ns}
\figalttext[Six plots in two rows, for FNS (top) and GNS (bottom), each showing median AUPR with interquartile range error bars for graph recovery, differential support and differential precision under the AND and OR symmetrisation rules.]{Six plots in two rows, for FNS (top) and GNS (bottom), each showing median AUPR with interquartile range error bars for graph recovery, differential support and differential precision under the AND and OR symmetrisation rules.}%
\end{figure}

\begin{figure}[h]%
\centering
\includegraphics[width=0.99\textwidth]{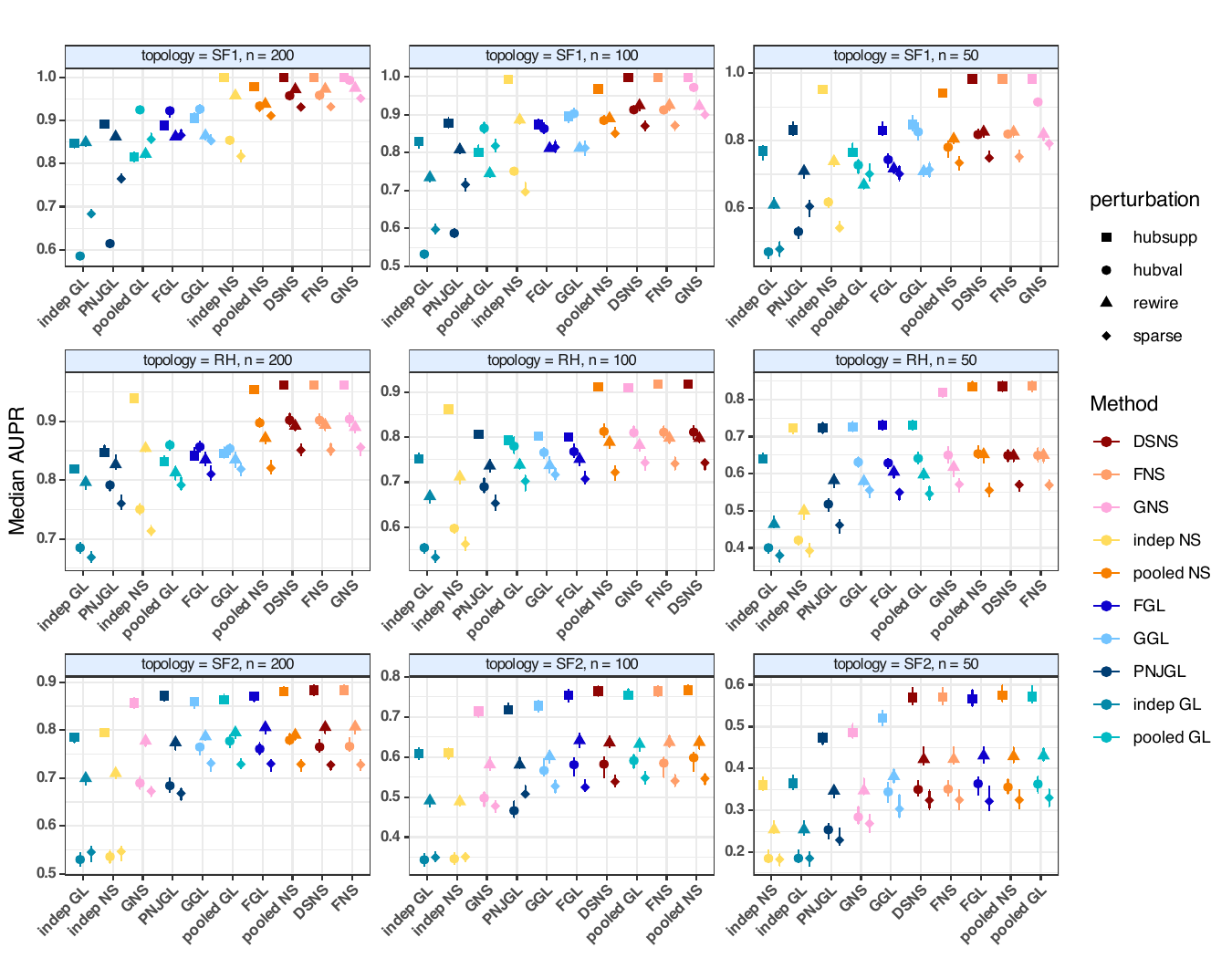}
\caption{Median area under the precision-recall curve for \textbf{graphs support recovery} across topologies, sample sizes and four perturbation types: 
sparse perturbations (\textit{sparse}), rewiring of edges (\textit{rewire}), 
removal of edges around hubs (\textit{hubsupp}), and differences of edge values connecting a hub (\textit{hubval}). Points represent median AUPR values and error bars indicate the interquartile range across simulation replicates. Methods are 
ordered by increasing median AUPR within each panel.}\label{suppfig_aupr_graph}
\figalttext[Grid of nine plots: three network topologies by three sample sizes, showing median AUPR.]{Grid of nine plots: three network topologies by three sample sizes (n = 200, 100, 50), showing median AUPR for ten methods with interquartile range error bars. Point shape distinguishes four perturbation types. Within each panel methods are ordered by increasing median AUPR.}%
\end{figure}

\begin{figure}[h]%
\centering
\includegraphics[width=0.99\textwidth]{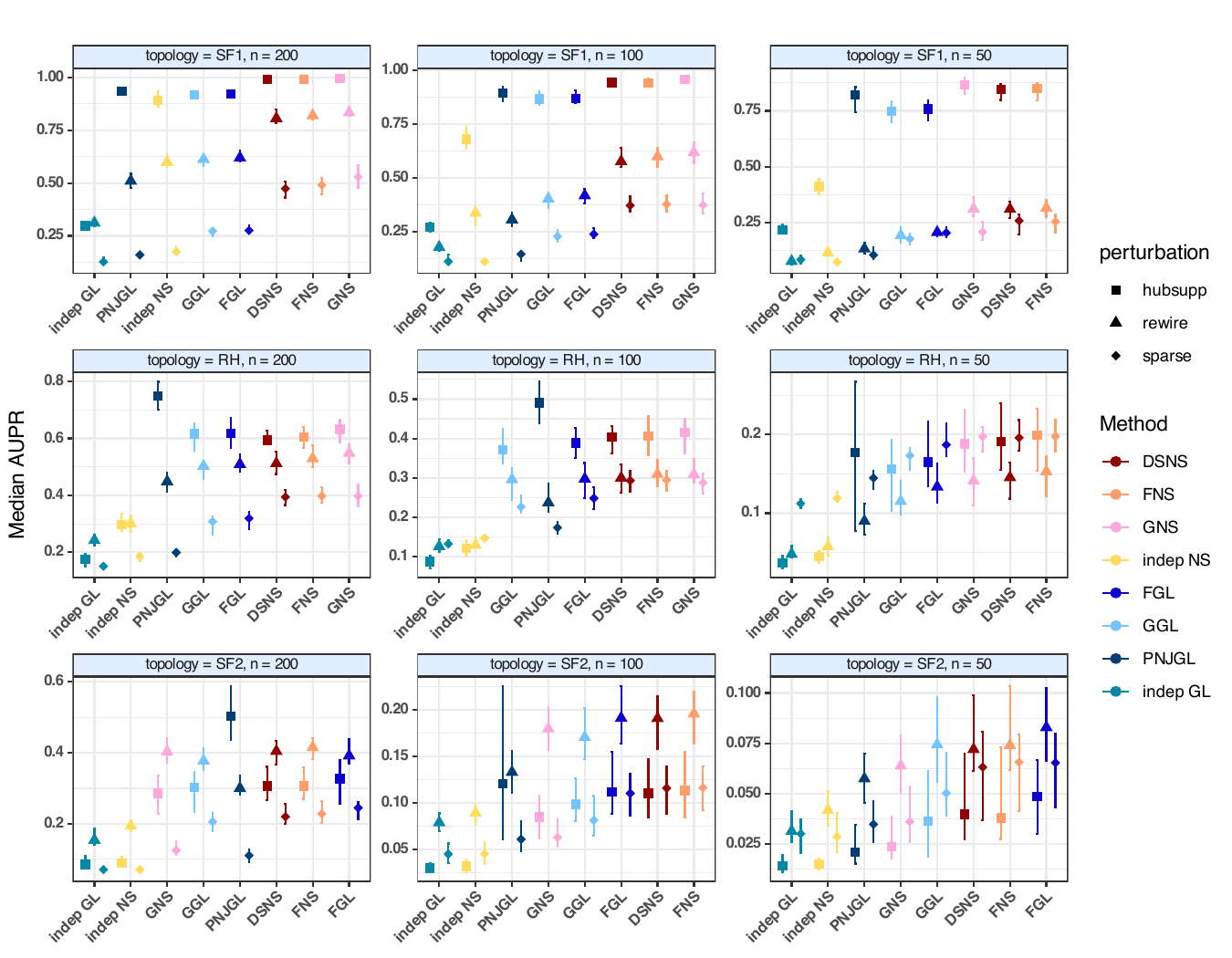}
\caption{Median area under the precision-recall curve for \textbf{differential support recovery} across graph topologies, sample sizes and four perturbation types: 
sparse perturbations (\textit{sparse}), rewiring of edges (\textit{rewire}) and 
removal of edges around hubs (\textit{hubsupp}). Points represent median AUPR values and error bars indicate the interquartile range across simulation replicates. Methods are 
ordered by increasing median AUPR within each panel.}\label{suppfig_aupr_suppdiff}
\figalttext[Grid of nine plots: three network topologies by three sample sizes, showing median AUPR.]{Grid of nine plots: three network topologies by three sample sizes, showing median AUPR for eight methods with interquartile range error bars, under three perturbation types distinguished by point shape.}%
\end{figure}

\begin{figure}[h]%
\centering
\includegraphics[width=0.99\textwidth]{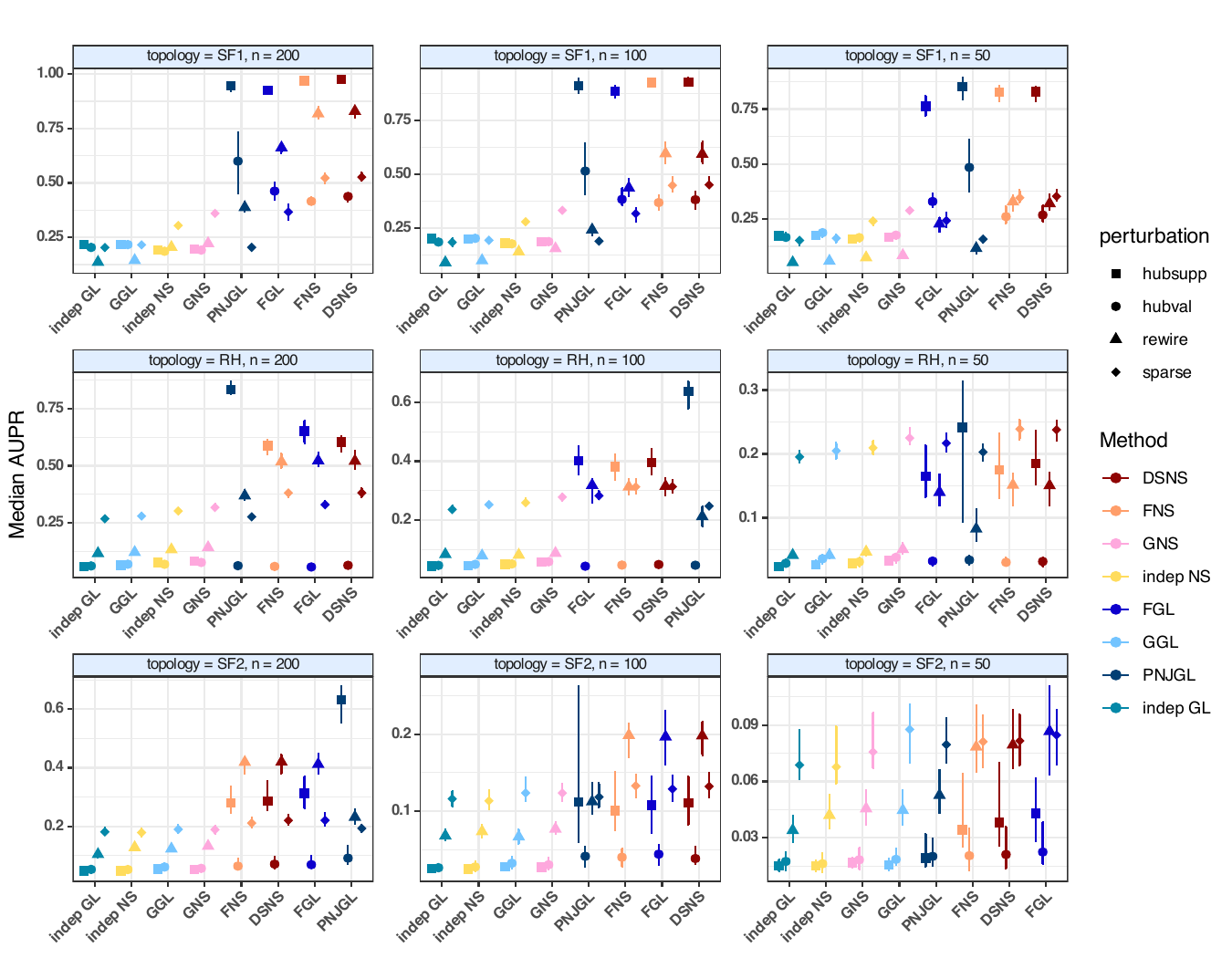}
\caption{Median area under the precision-recall curve for \textbf{differential precision edge recovery} across graph topologies, sample sizes and four perturbation types: 
sparse perturbations (\textit{sparse}), rewiring of edges (\textit{rewire}), 
removal of edges around hubs (\textit{hubsupp}) and differences of edge values connecting a hub (\textit{hubval}). Points represent median AUPR values and error bars indicate the interquartile range across simulation replicates. Methods are 
ordered by increasing median AUPR within each panel.}\label{suppfig_aupr_valdiff}
\figalttext[Grid of nine plots: three network topologies by three sample sizes, showing median AUPR.]{Grid of nine plots: three network topologies by three sample sizes, showing median AUPR for eight methods with interquartile range error bars, under four perturbation types distinguished by point shape.}
\end{figure}

\begin{figure}[h]%
\centering
\includegraphics[width=0.99\textwidth]{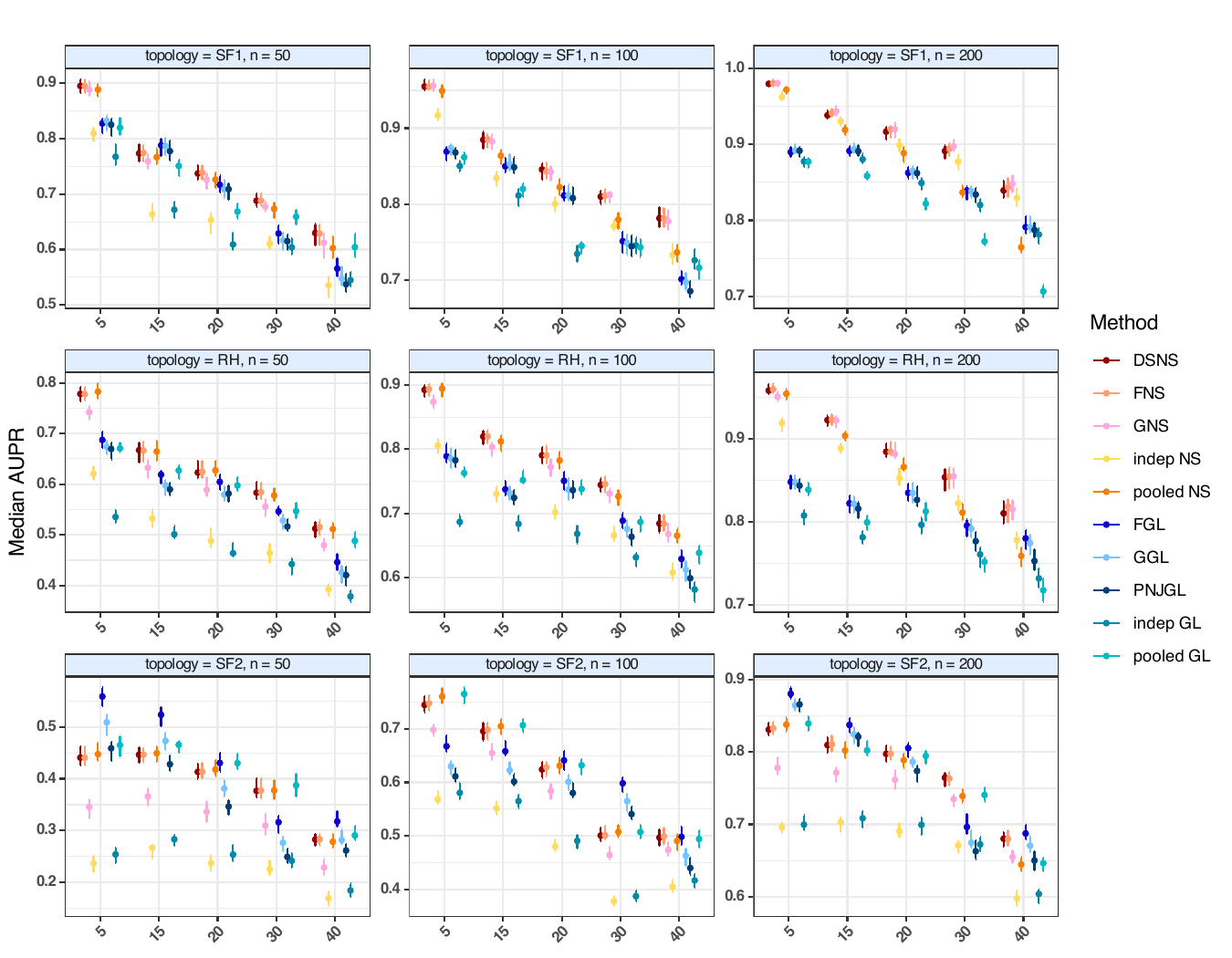}
\caption{Median area under the precision-recall curve for \textbf{graphs recovery} across topologies and sample sizes, for 5 different percentage of edges rewiring. Points represent median AUPR values and error bars indicate the interquartile range across simulation replicates.}
\label{suppfig_aupr_rewire_avg}
\figalttext[Grid of nine plots: three network topologies by three sample sizes, showing median AUPR.]{Grid of nine plots: three network topologies by three sample sizes, showing median AUPR of graphs recovery with interquartile range error bars for ten methods against the percentage of rewired edges, at 5, 15, 20, 30 and 40 per cent.}%
\end{figure}

\begin{figure}[h]%
\centering
\includegraphics[width=0.99\textwidth]{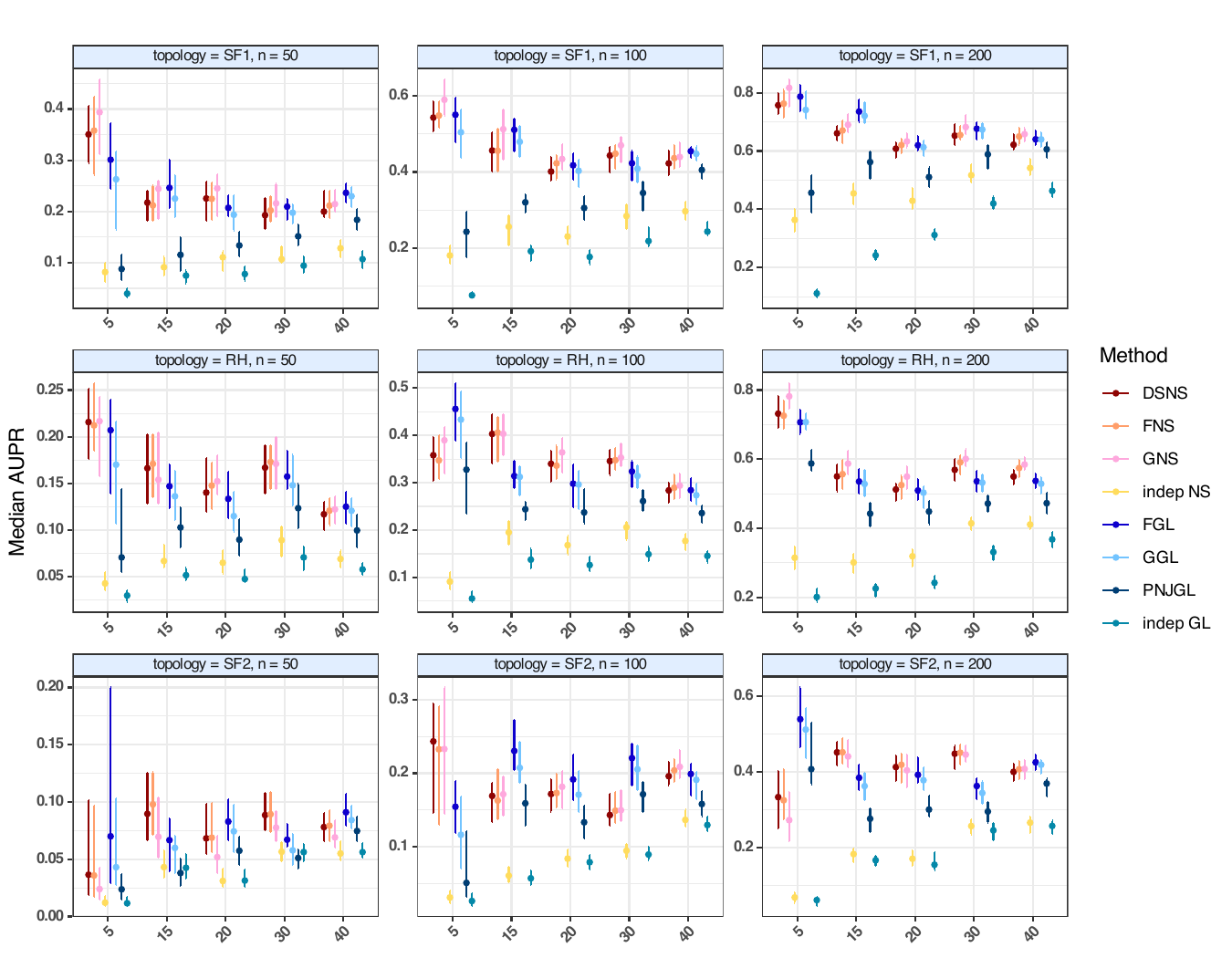}
\caption{Median area under the precision-recall curve for \textbf{differential support recovery} across topologies and sample sizes, for 5 different percentage of edges rewiring. Points represent median AUPR values and error bars indicate the interquartile range across simulation replicates.}\label{suppfig_aupr_rewire_suppdiff}
\figalttext[Grid of nine plots: three network topologies by three sample sizes, showing median AUPR.]{Grid of nine plots: three network topologies by three sample sizes, showing median AUPR of differential support recovery with interquartile range error bars for eight methods against the percentage of rewired edges, at 5, 15, 20, 30 and 40 per cent.}%
\end{figure}

\begin{figure}[h]%
\centering
\includegraphics[width=0.99 \textwidth]{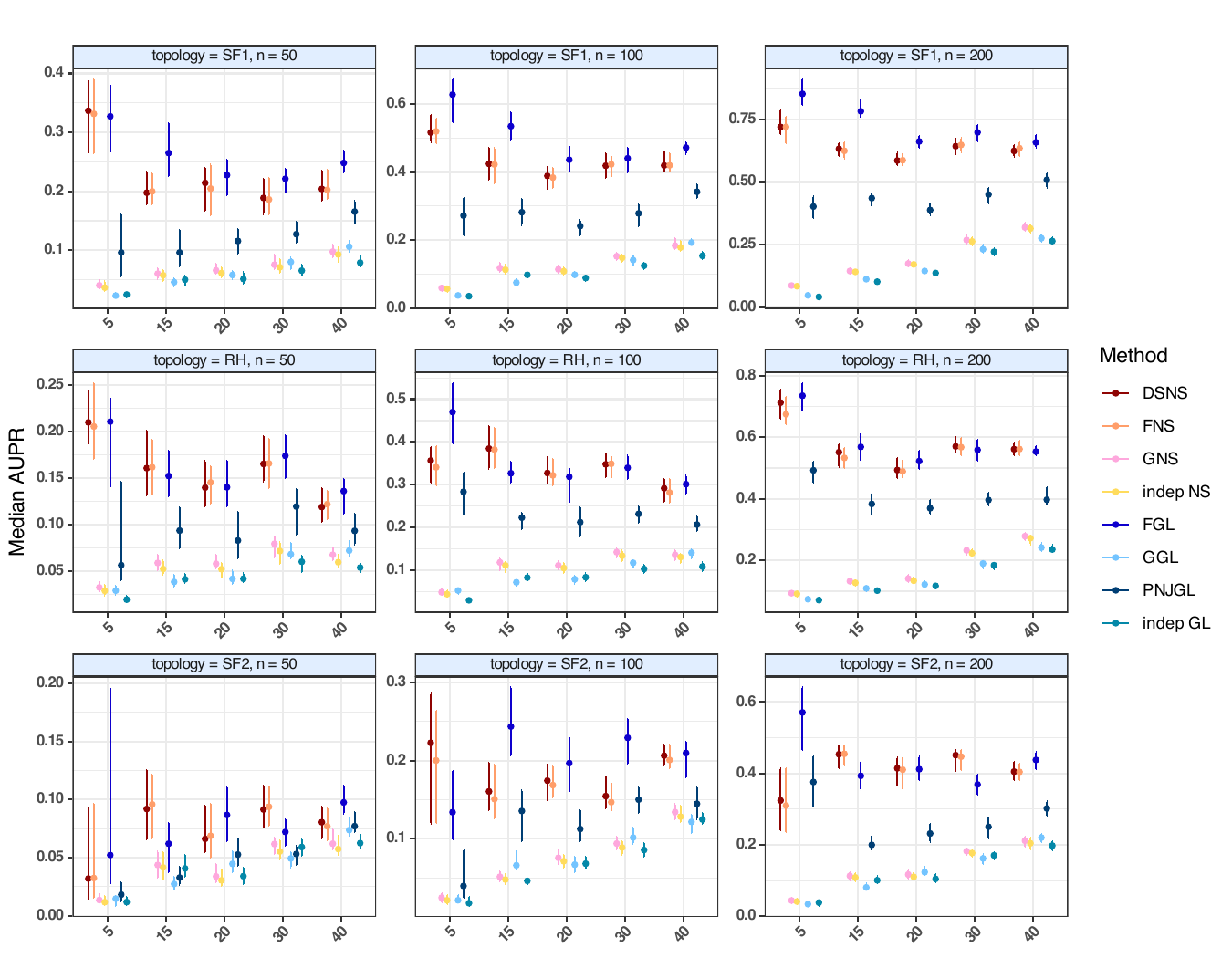}
\caption{Median area under the precision-recall curve for \textbf{differential precision edge recovery} across topologies and sample sizes, for 5 different percentage of edges rewiring. Points represent median AUPR values and error bars indicate the interquartile range across simulation replicates.}
\label{suppfig_aupr_rewire_valdiff}
\figalttext[Grid of nine plots: three network topologies by three sample sizes, showing median AUPR.]{Grid of nine plots: three network topologies by three sample sizes, showing median AUPR of differential precision edge recovery with interquartile range error bars for eight methods against the percentage of rewired edges, at 5, 15, 20, 30 and 40 per cent.}%
\end{figure}

\end{appendices}

\end{document}